\documentclass[12pt]{article}
\usepackage[toc,page]{appendix}
\usepackage{color, amsmath, amsthm}
 
\usepackage{graphicx}
\usepackage{amsfonts}
\usepackage{amssymb,amsmath,amscd}
\usepackage{fancyhdr}
\usepackage[top=2.5cm, bottom=2.5cm, left=2.5cm, right=2.5cm]{geometry} 
\usepackage{slashed}
\usepackage{hyperref}

\hypersetup{
    pdftitle={Rigid Supersymmetry on 5-Manifolds and Geometry},    
    pdfauthor={Yiwen Pan},     
    pdfnewwindow=true,      
    colorlinks=true,       
    linkcolor=black,          
    citecolor= black,        
    filecolor= black,      
    urlcolor= black           
}

\begin{document}

\title{Rigid Supersymmetry on 5-dimensional Riemannian Manifolds and Contact Geometry}
\date{}
\author{Yiwen Pan\footnote{Email address:  \href{mailto:yiwen.pan@stonybrook.edu}{yiwen.pan@stonybrook.edu}}
\\ \textit{C.N. Yang Institute for Theoretical Physics, Stony Brook, NY, 11790, USA}}
\maketitle
\begin{abstract}
	\noindent In this note we generalize the methods of \cite{Dumitrescu:2012ys}\cite{Dumitrescu:2012ly}\cite{Closset:2012zr} to 5-dimensional Riemannian manifolds $M$. We study the relations between the geometry of $M$ and the number of solutions to a generalized Killing spinor equation obtained from a 5-dimensional supergravity. The existence of 1 pair of solutions is related to almost contact metric structures. We also discuss special cases related to $M = S^1 \times M_4$, which leads to $M$ being foliated by submanifolds with special properties, such as Quaternion-K\"ahler. When there are 2 pairs of solutions, the closure of the isometry sub-algebra generated by the solutions requires $M$ to be $S^3$ or $T^3$-fibration over a Riemann surface. 4 pairs of solutions pin down the geometry of $M$ to very few possibilities. Finally, we propose a new supersymmetric theory for $\mathcal{N} = 1$ vector multiplet on K-contact manifold admitting solutions to the Killing spinor equation.
\end{abstract}

\thispagestyle{fancy}



\newpage
\tableofcontents

\numberwithin{equation}{section}

\section{Introduction}

As discussed in \cite{Festuccia:2011ve}, to obtain a supersymmetric theory on a Riemannian manifold $M$, one can first couple the desired multiplet to supergravity, then take the rigid limit, sending the Planck mass to infinite. 

In the process of taking the limit, one keeps auxiliary fields instead of imposing their equations of motion. If a background of auxiliary fields and metric is invariant under the supergravity transformation, they actually give rise to rigid supersymmetry.

This line of reasoning has been utilized to study $4d$ $\mathcal{N}=1$ supersymmetry with/without $U(1)_R$ symmetry \cite{Dumitrescu:2012ys}\cite{Dumitrescu:2012ly}\cite{Klare:2012aa}, and $3d$ $\mathcal{N}=2$ supersymmetry with $U (1)_R  $\cite{Closset:2012zr}. In these papers, the existence of a number of supercharges is proven to be related to the geometric structure of $M$. For instance, on any $4d$ Hermitian manifold there exists at least one supercharge\cite{Dumitrescu:2012ys}\cite{Klare:2012aa}, and $3d$ manifolds with an almost contact structure admit at least one supercharge. Similar discussions for manifolds with Lorentz signatures in dimension 3 and 4 can be found in \cite{Hristov:2013aa} \cite{Cassani:2012aa}.

In 5-dimension, there are rapidly growing literatures on constructing 5d supersymmetric theories, as well as their relations with 6d (2,0) theories and lower dimensional Chern-Simons theories.

For example, in \cite{Hosomichi:2012fk}, a supersymmetric gauge theory on $S^5$ is obtained from 5d supergravity, with the well-known Killing spinor equation
	\begin{equation}
		{\nabla _m}{\xi _I} = {t_I}^J{\Gamma _m}{\xi _J}.
		\label{Killing-Hosomichi}
	\end{equation}
Using the supersymmetry algebra, the author proposed adding a term $\delta {\rm{tr}}( {{{\left( {\delta \lambda } \right)}^\dag }\lambda } )$ to the Lagrangian, and derived the localization condition. The localization condition is further used in \cite{Kallen:2012fk}\cite{Kallen:2012zr} to analyze physical and twisted supersymmetric gauge theories coupled with matter defined on a principal $U(1)$ bundle $M_5$ over a 4-dimensional manifold. In particular, the perturbative partition function on $S^5$ is computed. Their localization result leads to the derivation of $N^3$-behavior of the free energy of 5d SYM, in the large 't Hooft coupling limit\cite{Kallen:2012dq}. Complete localization of the partition function on $S^5$ is carried out in \cite{Kim:2012aa}\cite{Kim:2012uq}. The authors first computed the perturbative contribution and conjectured the non-perturbative contribution by requiring the full partition function to be identical to a 6d index \cite{Kim:2012aa}. In their subsequent work \cite{Kim:2012uq} the full partition function is computed which proves the conjecture.

There are also supersymmetric theories constructed by hand or by dimensional reduction from 6d.
In \cite{Terashima:2012qf} supersymmetric gauge theory on $S^1 \times S^4$ is obtained, and the localization is carried out. \cite{Cordova:2013kx} derived a class of 5d SYM theories from 6d (2,0) supergravity. Further in \cite{Cordova:2013vn}\cite{Lee:2013aa}\cite{Yagi:2013aa}, supersymmetric theories on $S^3 \times M_3$ obtained from M5-brane are shown to be equivalent to 3d complexified Chern-Simons theory. Supersymmetric theories on $\mathbb{C}P^2 \times \mathbb{R}$ were also obtained from 6d and studied in details in \cite{Kim:2012ab}\cite{Kim:2013aa}.

A complete picture of the relation between supersymmetry and geometry of $M$, however, is not clear. It is reasonable to believe that methods similar to those in \cite{Closset:2012zr} can be straight-forwardly applied to 5-dimensional manifolds.

In this paper, we take a step towards such understanding, and as expected, the results turn out to be closely related to contact and almost contact structures on 5-manifolds.

We use the minimal off-shell 5d Supergravity discussed in \cite{Zucker:2000uq} and focus on the Killing spinor equation (\ref{Killing-equation})
	\begin{equation}
		{\nabla _m}{\xi _I} = {t_I}^J{\Gamma _m}{\xi _J} + \frac{1}{4}{V^{pq}}{\Gamma _{mpq}}{\xi _I} + \frac{1}{2}{F_{mn}}{\Gamma ^n}{\xi _I} + {\left( {{A_m}} \right)_I}^J{\xi _J},
	\end{equation}
coming from requiring supergravity variation $\delta \psi_m^I$ of gravitino $\psi$ to vanish. We study many interesting necessary conditions for $M$ to admit different number of solutions to the Killing spinor equation, by partially solving the auxiliary fields in terms of the bilinears, and discussing special cases which are related to well-known geometries and results in lower dimension supersymmetry. However, the results we obtain are not a complete classification of manifolds admitting solutions.

In the end, we propose a 5-dimensional supersymmetric theory for the $\mathcal{N} = 1$ vector multiplet, which can be defined on a contact manifold with an associated metric admitting solution to equation (\ref{SUSY-Killing}). However, it should be pointed out that this theory is not obtained directly from supergravity, since we started from 5-dimensional Weyl multiplet without coupling to matter. Therefore the rigid limit of the supergravity action on a fixed background gives a number rather than a supersymmetric theory. To obtain the final supersymmetric background, one also needs to require the background auxiliary fields to satisfy a more complicated differential equation from the vanishing of the supergravity variation of another spinor field in the Weyl multiplet. In this sense, the present work covers an important sector of the problem, but a complete analysis requires further exploration.

	\vspace{20pt}
	This paper is organized as follows:
	\begin{itemize}
	\item In section \ref{section-2}, we briefly review Zucker's 5d $\mathcal{N} = 1$\footnote{It is called ``$\mathcal{N} = 2$" in his paper, but it really means a theory with 8 supercharges, which supersymmetry parameter an $SU(2)$-symplectic Majorana spinor $\xi_I^\alpha$.} minimal supergravity and the Killing spinor equation from the vanishing gravitino variation.
	
	\item In section 3, we study the bilinears constructed from one or two symplectic Majorana spinors. We see that when a global nowhere-vanishing section of $ad(P_{SU(2)})$ over $M$ exists, $M$ has an {almost contact structure} corresponding to each spinor.
	
	\item In section 4, we start with a general discussion of the Killing spinor equation (\ref{Killing-equation})
	\begin{equation}
		{\nabla _m}{\xi _I} = {t_I}^J{\xi _J} + \frac{1}{2}{V^{pq}}{\Gamma _{mpq}}{\xi _I} + \frac{1}{2}{F_{mn}}{\Gamma ^n}{\xi _I} + {\left( {{A_m}} \right)_I}^J{\xi _J},
	\end{equation}
including its shifting symmetry and Weyl symmetry. We then analyze the necessary conditions on the geometry of $M$ such that it admits certain number of solutions. For each (pair of) solutions, we see that the auxiliary fields can be partially solved in terms of the bilinears, and the Killing spinor equation is then simplified using a compatible connection $\hat \nabla$
	\begin{equation}
		\boxed{{{\hat \nabla }_m}{{\hat \xi }_I} - {({{\hat A}_m})_I}^J{{\hat \xi }_J} = 0}
	\end{equation}
Some special cases related to product form $M = S^1 \times M_4$ are discussed: depending on the field configuration, one obtains geometrical restriction of $M_4$ being K\"ahler, Quaternion K\"ahler or HyperK\"ahler,  or a reduction of our 5d Killing spinor equation to 4d cases discussed in literatures\cite{Dumitrescu:2012ys}\cite{Dumitrescu:2012ly}. For 2 (pairs of ) supercharges to exist, we will see that the geometry of $M$ is heavily constrained by the isometry algebra to be $T^3$ or $S^3$-fibration over Riemann surface $\Sigma$.

For 4 pairs of supercharges to exist, we will see that there are only 3 possible cases, which basically fixes the geometry of $M$.

	\item In section 5, we propose a new supersymmetric theory for the $\mathcal{N} =1$ vector multiplet, which can be defined on K-contact manifolds $(M,g, \kappa)$ admitting solutions to equation
	\begin{equation}
		{D _m}{\xi _I} = {t_I}^J{\xi _J} + \frac{1}{4}{{{\cal F}}^{pq}}{\Gamma _{mpq}}{\xi _I} + \frac{1}{2}{{{\cal F}}_{mn}}{\Gamma ^n}{\xi _I}
	\end{equation}
with $\mathcal{F}$ any ``anti-self-dual" (defined later) closed 2-form.

	\item In section 7, we provide a few examples of solving Killing spinor equations on selected manifolds to illustrate some results obtained in previous sections.
	
	\item In the appendix, we review conventions on gamma matrices and differential geometry as well as necessary mathematical backgrounds on contact geometry. Useful formula are also listed.
	
	\end{itemize}

\section{$\mathcal{N} = 1$ Minimal Off-shell Supergravity}\label{section-2}

5 dimensional minimal off-shell supergravity was studied by Zucker \cite{Zucker:2000uq}\footnote{It is called $\mathcal{N} = 2$ in \cite{Zucker:2000uq}, however, it actually has 8 supercharges following from the symplectic Majorana reality condition and it is more sensible to call it $\mathcal{N} =1$}. In his paper, the linearized gravity multiplet and its SUSY transformation is obtained through coupling to the current multiplet of supersymmetric Maxwell multiplet. Then the linearized multiplet is covariantized (making the transformation local) and its supergravity transformation can be derived. In this section we summarize his work, and obtain the Killing spinor equation needed for the rigid limit.

The super-Maxwell multiplet consists of the field content $(\varphi, A, \lambda')$, where $\varphi$ is a real scalar, $A$ is a gravi-photon with field strength $f_{mn} = {\partial _m}{A_n} - {\partial _n}{A_m}$, and $\lambda'$ is the gaugino.

The Lagrangian reads
	\begin{equation}
		\mathcal{L} =  - \frac{1}{4}{f_{mn}}{f^{mn}} + \frac{1}{2}{\partial _m}\varphi {\partial ^m}\phi  + \frac{i}{2}\bar \lambda' {\Gamma ^m}{\partial _m}\lambda'.
	\end{equation}
	
	The Lagrangian is invariant under the on-shell supersymmetry transformation
	\begin{equation}
		\delta \varphi  = i\bar \epsilon \lambda' ,\;\delta {A_m} = i\bar \epsilon {\Gamma _m}\lambda' ,\;\delta \lambda'  = \frac{1}{2}{f_{mn}}{\Gamma ^{mn}}\epsilon  - {\partial _m}\phi {\Gamma ^m}\epsilon ,
	\end{equation}
which form a closed algebra modulo the equation of motion:
	\begin{equation}
		{\Gamma ^m}{\partial _m}\lambda'  = 0.
	\end{equation}

	There are several symmetries of the theory:
	\begin{itemize}
	\item Spacetime symmetry, whose conserved current is the energy-momentum tensor
		\begin{equation}
			{T_{mn}} =  - {f_{mk}}{f_n}^k + \frac{1}{2}{\eta _{mn}}{f_{kl}}{f^{kl}} + {\partial _m}\varphi {\partial _n}\varphi  - \frac{1}{2}{\eta _{mn}}{\left( {\partial \varphi } \right)^2} + \frac{i}{4}\bar \lambda' \left( {{\Gamma _m}{\partial _n} + {\Gamma _n}{\partial _m}} \right)\lambda' .
		\end{equation}
	
	\item Supersymmetry, whose the conserved current is
		\begin{equation}
			{J^m} = {\Gamma ^n}{\Gamma ^m}\lambda' {\partial _m}\varphi  + \frac{1}{2}{f_{nl}}{\Gamma ^{nl}}{\Gamma ^m}\lambda' .
		\end{equation}
	
	\item $SU(2)$ $R$-symmetry, whose the conserved $R$-current is
		\begin{equation}
			J_m^a = \bar \lambda' {\tau ^a}{\Gamma _m}\lambda' .
		\end{equation}

	\end{itemize}
	
	These currents can form a supermultiplet if proper additional objects are added to close the algebra. The complete current multiplet consists of 
	\begin{equation}
		\left( {C,\zeta ,{X^a},{w_{mn}},J_m^a,{J_m},{j^a},{T^{mn}}} \right).
	\end{equation}

	Then one can couple this multiplet to linearized gravity. The bosonic components of the multiplet are $\left( {{h_{mn}},{V_{mn}},{a_m},t,C} \right)$, where $a_m$ is $U(1)$ gauge field with field strength $F_{mn} = {\partial _m}{a_n} - {\partial _n}{a_m}$. The Fermions are an auxiliary spinor $\lambda$ of dimension 3 (not to be confused with the gaugino $\lambda_I$ of the $\mathcal{N} =1$ vector multiplet in a later section) and the gravitino $\psi^m_I$
	\begin{equation}
		\mathcal{L} = \frac{1}{8}{h_{mn}}{T^{mn}} + \frac{i}{4}{\bar J^m}{\psi _m} - 4C'C - 2i\bar \zeta \lambda  - \frac{1}{2}{w_{mn}}{V^{mn}} + {X^a}{t^a} + \frac{1}{{2\sqrt 3 }}{A_m}J_{\left( 1 \right)}^m + \frac{1}{4}J_m^aV_a^m.
	\end{equation}

	Requiring the Lagrangian to be supersymmetric, one obtains supergravity transformation (with parameter $\xi_I$ which is a symplectic Majorana spinor) of the linearized multiplet. Further covariantizing the transformation gives the full Supergravity transformation (here we only list schematically first few lines and omit coefficients in front of each term)
	\begin{equation}
		\left\{ \begin{array}{l}
		\delta e_m^i\sim{\xi _I}{\Gamma ^i}\psi _m^I\\[0.8em]
		\delta {A_m}\sim{\xi _I}{\psi _m}\\[0.8em]
		\delta {\psi _m}\sim \mathcal{D}_m^{\hat \omega }{\xi _I} + {{\hat F}_{mn}}{\Gamma ^n}{\xi _I} + {V^{pq}}{\Gamma _{mpq}}{\xi _I} + {\left( {{A_m}} \right)_I}^J{\xi _J} + ...\\[0.8em]
		\delta \lambda_I = \nabla_m V^{mn} \Gamma_n \xi_I + *(V \wedge V)^m \Gamma_m \xi_I + V_{mnk}\Gamma^{mnk} \xi_I + ...
		\end{array} \right.,
	\end{equation}
where $...$ in the third line denotes terms that will vanish when taking rigid limit. In the last line we schematically show a few terms involving $V$, and use $...$ to denote remaining complicated terms. 

	The rigid limit procedure sets fermions to zero, keeping only the bosonic fields (metric and other fields) to some background which needs to be determined. If such background is invariant under certain supergravity transformation, in particular, $\delta \psi  = 0$, one obtain a rigid supersymmetric background with the resulting metric. 
	
	The condition $\delta \psi = 0$ reads, with some coefficients reinstalled without loss of generality,
	\begin{equation}
		\delta {\psi _{mI}} = {\nabla _m}{\xi _I} - {t_I}^J{\Gamma _m}{\xi _J} - \frac{1}{2}{F_{mn}}{\Gamma ^n}{\xi _I} - \frac{1}{2}{V^{pq}}{\Gamma _{mpq}}{\xi _I} - {\left( {{A_m}} \right)_I}^J{\xi _J} = 0
	\label{Killing-equation}.
	\end{equation}
which is the Killing spinor equation we are going to analyze in the following sections.

	In principle one needs to also solve the equation from $\delta \lambda = 0$ in taking the rigid limit. However, in this paper we do not discuss this equation, but rather focus on the simpler yet important Killing spinor equation (\ref{Killing-equation}).

\section{Symplectic Majorana spinor and bilinears}

	In this section, we review the properties of symplectic Majorana spinor and their bilinears. Note that we consider bosonic spinors in the following discussions.
	
	\vspace{20pt}
	
	On a 5-dimensional Riemannian manifold $M$, one can define Hermitian Gamma matrices, the charge conjugation matrix and $SU(2)$ symplectic Majorana spinors\footnote{Note that ordinary Majorana condition cannot be defined in 5d.}. 
	
	Hermitian Gamma matrices are denoted as $\Gamma$
	\begin{equation}
		\left\{ {{\Gamma _m},{\Gamma _n}} \right\} = 2{g_{mn}},
	\end{equation}
and hermiticity implies
	\begin{equation}
		\overline {{\Gamma _m}}  = {\left( {{\Gamma _m}} \right)^T}.
	\end{equation}
	
	The Charge conjugation matrix is denoted as $C$,
	\begin{equation}
		C{\Gamma ^m}{C^{ - 1}} = {\left( {{\Gamma ^m}} \right)^T} = \overline {{\Gamma _m}} .
	\end{equation}
	
	We also define the $SU(2)$-invariant tensor $\epsilon^{IJ}$ and $\epsilon_{IJ}$
	\begin{equation}
		{\epsilon ^{12}} =  - {\epsilon ^{21}} =  - {\epsilon _{12}} = 1,
	\end{equation}
and raising and lowering convention
	\begin{equation}
		{\epsilon _{IJ}}{X^J} = {X_I},\;{\epsilon ^{IJ}}{X_J} = {X^I}.
	\end{equation}
	
	With these quantities we define the symplectic Majorana spinor condition as
	\begin{equation}
		\overline {\xi _I^\alpha }  = {\epsilon ^{IJ}}{C_{\alpha \beta }}\xi _J^\beta ,
	\end{equation}
and a $\mathbb{C}$-valued inner product of any two spinors denoted by parenthesis $()$
	\begin{equation}
		\left( {\xi \eta } \right) \equiv {\xi ^\alpha }{C_{\alpha \beta }}{\eta ^\beta },
	\end{equation}
and further a positive-definite inner product $(\; ,\;)$ between symplectic Majorana spinors $\xi,\eta$
	\begin{equation}
		\left( {\xi ,\eta } \right) \equiv {\epsilon ^{IJ}}\left( {{\xi _I}{\eta _J}} \right).
	\end{equation}
	
\subsection{Bilinears from 1 symplectic Majorana spinor}

	Now we're ready to define bilinears constructed from one symplectic Majorana spinor $\xi_I$.
	
		(1) Function $s\in C^{\infty} (M)$:
	\begin{equation}
		s \equiv {\epsilon ^{IJ}}\left( {{\xi _I}{\xi _J}} \right) = 2\left( {{\xi _1}{\xi _2}} \right).
	\end{equation}

	Note that this function is strictly positive if $\xi$ is nowhere-vanishing:
	\begin{equation}
		s = {\epsilon ^{IJ}}\xi _I^\alpha {C_{\alpha \beta }}\xi _J^\beta  = \sum\limits_\alpha  {\xi _I^\alpha \overline {\xi _I^\alpha } }  > 0.
	\end{equation}

	(2) Vector field $R \in \Gamma (TM)$:
	\begin{equation}
		{R_m} \equiv {\epsilon ^{IJ}}{\xi _I}{\Gamma _m}{\xi _J},
	\end{equation}
and the corresponding 1-form
	\begin{equation}
		\kappa_m \equiv g_{mn}R^n,
	\end{equation}
which implies,  when acting on $\Omega^p(M)$
	\begin{equation}
		{\iota _R} \circ * = {\left( { - 1} \right)^p}* \circ \left( {\kappa  \wedge } \right).
	\end{equation}
	
	(3) 2-form\footnote{One could of course go on defining higher forms $\Theta _{lmn}^{IJ} \equiv {\xi ^I}{\Gamma _{lmn}}{\xi ^J}$ and $\Theta _{mnpq}^{IJ} \equiv {\xi ^I}{\Gamma _{mnpq}}{\xi ^J}$, but duality of Gamma matrices gives
	\begin{equation}
		\Theta _{lmn}^{IJ} =  - \frac{{\sqrt g }}{{2!}}{\epsilon ^{pq}}_{lmn}\Theta _{pq}^{IJ},
	\end{equation}
	and
	\begin{equation}
		\Theta _{mnpq}^{IJ} = \sqrt g {\epsilon ^r}_{mnpq}\Theta _r^{IJ},
	\end{equation}
	}
	\begin{equation}
		\Theta _{mn}^{IJ} \equiv \left( {{\xi ^I}{\Gamma _{mn}}{\xi ^J}} \right),
	\end{equation}
with symmetry
	\begin{equation}
		{\Theta ^{IJ}} = {\Theta ^{JI}}.
	\end{equation}

	Let  $t_{IJ}$ be an arbitrary triplet of functions, namely
	\begin{equation}
		t_{IJ} = t_{JI}, I = 1,2;
	\end{equation} 
then its contraction with $\Theta$ gives a real 2-form
	\begin{equation}
		\left( {t \Theta } \right) \equiv {t ^I}_J{\left( {{\Theta ^J}_I} \right)}.
	\end{equation}
	
	Using the Fierz identities one can derive useful relations between these quantities, which we list in appendix \ref{Appendix-E}. 

	\vspace{20pt}
	Given the nowhere-vanishing 1-form $\kappa$ and the vector field $R$, one can decompose the tangent bundle $TM = TM_H \oplus TM_V$, where at any point $p\in M$, $TM_H |_p$ is annihilated by $\kappa$, while $TM_V$ is a trivial line bundle generated by $R$. Let's call $TM_H$, and similarly all tensors annihilated by $\kappa$ (or $R$) ``horizontal", while those in the orthogonal complement "vertical". In particular, one has decompositions
	\begin{equation}
		{\Omega ^2}\left( M \right) = \Omega _V^2\left( M \right) \oplus \Omega _H^2\left( M \right) = \kappa  \wedge \Omega _H^1\left( M \right) \oplus \Omega _H^2\left( M \right)
	\end{equation}

	For an arbitrary nowhere-vanishing triplet of functions $t_{IJ}$ with the property (readers may find conventions in Appendix \ref{Appendix-B})
	\begin{equation}
		t_{IJ} = t_{JI}, \; \; \overline {{t_{IJ}}}  = {\epsilon ^{II'}}{\epsilon ^{JJ'}}{t_{I'J'}}
	\end{equation}
one can define a map ${\varphi _t }:\Gamma \left( {TM} \right) \to \Gamma \left( {TM} \right)$ as
	\begin{equation}
		{\left( {{\varphi _t }} \right)_m}^n \equiv \frac{1}{s}\sqrt {\frac{{ - 2}}{{{\rm{tr}}\left( {{t ^2}} \right)}}} {\left( {t \Theta } \right)_m}^n.
	\end{equation}
Obviously, one has
	\begin{equation}
		{\varphi _t } \circ {\varphi _t } =  - 1 + {s^{ - 2}}R \otimes \kappa ,
	\end{equation}
and when restricted on $T M_H$, $\varphi_t$ is some sort of a``complex" structure:
	\begin{equation}
		{\left. {{\varphi _t } \circ {\varphi _t }} \right|_{T{M_H}}} =  - 1.
	\end{equation}
	
	Together with the vector field $s^{-1}R$ and 1-form $s^{-1}\kappa$, $\varphi_\lambda$ defines an {almost contact structure} on $M$\cite{Blair} (see also Appendix \ref{Appendix-D}).
	
	Finally, let us comment on the ``(anti)self-dual" horizontal forms. Define operator ${*_H} \equiv s^{-1}{\iota _R}*$, which is the hodge dual ``within" horizontal hyperplanes. It is easy to verify that acting on any horizontal $p$-forms
	\begin{equation}
		*_H^2 = {\left( { - 1} \right)^p}.
	\end{equation}
In particular, we decompose the horizontal 2-forms into 2 subspaces according to their eigenvalues of $*_H$
	\begin{equation}
		\Omega _H^2 = \Omega _H^{2 + } \oplus \Omega _H^{2 - }, \;\; {*_H}\omega _H^ \pm  =  \pm \omega _H^ \pm, \forall \omega_H^\pm \in \Omega_H^\pm.
		\label{self-duality}
	\end{equation}
We call the horizontal forms in $\Omega _H^{2 + }$ ``{self-dual}", while the others ``{anti-self-dual}". Clearly, these 2 notions are interchanged as one flips the sign of the vector field $R$, hence this notion of ``self-duality" is not as intrinsic as the well-established notion of self-duality on 4-dimensional oriented manifolds. 
	
	Suppose $\Omega^+$ is a self-dual 2-form. Then it satisfies, by definition, 
	\begin{equation}
		\frac{{\sqrt g }}{2s}{\epsilon ^{pq}}_{lmn}{R^l}{\Omega^+ _{pq}} = {\Omega ^+_{mn}}.
	\end{equation}
	
	It follows immediately that
	\begin{equation}
		\boxed{
			{\Omega ^+_{mn}}{\Gamma ^{mn}}{\xi _I} = 0
			},
	\end{equation}
using the fact that the inner product $\left( {\psi ,\psi } \right) \equiv {\epsilon ^{IJ}}\left( {{\psi _I}{\psi _J}} \right)$ is positive definite, and the action of $\Gamma_{mn}$ preserve symplectic Majorana property.

\subsection{Bilinears from 2 symplectic Majorana spinors}

	In this section, we consider the case when there are 2 symplectic Majorana spinors, and analyze their bilinears.
	
	Denote the two spinors $\xi_I$ and $\tilde \xi_I$. Obviously they each generates a set of quantities as we discussed in the previous sections: $(s,R, \kappa, \Theta)$ and $(\tilde s, \tilde R, \tilde \kappa, \tilde \Theta)$.
	
	In addition to these quantities, they form some new mixed bilinears. Conventions for $IJ$ indices can be found in appendix \ref{Appendix-B}.
	
	\begin{itemize}
		\item Functions
		\begin{equation}
			{u_{IJ}} \equiv ( {{\xi _I}{\tilde \xi _J}} ),
		\end{equation}
	with triplet-singlet decomposition
		\begin{equation}
			{u_{IJ}} = {u_{\left( {IJ} \right)}} + {u_{\left[ {IJ} \right]}} = { \hat u_{IJ} } - \frac{1}{2}{\epsilon _{IJ}}u,
		\end{equation}
	where
		\begin{equation}
			u \equiv {\epsilon ^{IJ}}{u_{IJ}}.
		\end{equation}
		
		Notice that
		\begin{equation}
			\overline {{u_{IJ}}}  =   {\epsilon ^{II'}}{\epsilon ^{JJ'}}{u_{I'J'}} \equiv   {u^{IJ}},
		\end{equation}
	and in particular function $u$ is real-valued
		\begin{equation}
			\overline u  = u = \sum\limits_I {\xi _I^\alpha \overline {\tilde \xi _I^\alpha } } ,
		\end{equation}
	which results in positivity
		\begin{equation}
			{u_{IJ}}{u^{IJ}} = \sum {{u_{IJ}}\overline {{u_{IJ}}} } = \frac{1}{2} u^2 + \hat u_{IJ} \hat u ^{IJ}\ge 0.
		\end{equation}

		\item Vector fields $Q_{IJ}$
		\begin{equation}
			Q_{IJ}^m \equiv ( {{\xi _I}{\Gamma ^m}{\tilde \xi _J}} ),
		\end{equation}
	with a decomposition
		\begin{equation}
			{Q_{IJ}} = \hat Q_{IJ} - \frac{1}{2}{\epsilon _{IJ}}Q,
		\end{equation}
	where
		\begin{equation}
			{Q^m} \equiv {\epsilon ^{IJ}}( {{\xi _I}{\Gamma ^m}{\tilde \xi _J}} ).
		\end{equation}
		
		Note that similar to the function case, we have
		\begin{equation}
			\overline {{Q_{IJ}}}  =  {Q^{IJ}},
		\end{equation}
	and in particular a real vector field
		\begin{equation}
			\overline{Q} = Q.
		\end{equation}
		
		We denote corresponding 1-forms 
		\begin{equation}
			\displaystyle {\tau _{IJ}} \equiv {\left( {{Q_{IJ}}} \right)_m}d{x^m} = {\hat \tau _{IJ}} - \frac{1}{2}{\epsilon _{IJ}}\tau.
		\end{equation}
				
		\item Two forms
		\begin{equation}
			\chi _{mn}^{IJ} \equiv ( {{\xi ^I}{\Gamma _{mn}}{\tilde \xi ^J}} ).
		\end{equation}
		
		Also we define
	\begin{equation}
		\chi  \equiv {\epsilon ^{IJ}}{\chi _{IJ}},\;\;{{\hat \chi }_{IJ}} = {\chi _{\left( {IJ} \right)}}.
	\end{equation} 
	
	\end{itemize}

	\vspace{10pt}
	
	These bilinears satisfy various algebraic relations. Here we list some relevant formulas.
	\begin{itemize}
	
		\item \textbf{Norms and inner products of vector fields}
		
		(1)
		\begin{equation}
			R \cdot \tilde R = 4{u_{IJ}}{u^{IJ}} - s\tilde s \Rightarrow \left\{ \begin{array}{l}
		{\left| {\tilde sR + s\tilde R} \right|^2} = 8s\tilde s{u_{IJ}}{u^{IJ}}\\[0.5em]
		{\left| {\tilde sR - s\tilde R} \right|^2} = 4s\tilde s\left( {s\tilde s - 2{u_{IJ}}{u^{IJ}}} \right)
\end{array} \right.
		\end{equation}
		
		(2)
		\begin{equation}
			{Q_{IJ}} \cdot {Q_{KL}} = 2{u_{IL}}{u_{KJ}} - {u_{IJ}}{u_{KL}} - \frac{1}{2}{\epsilon _{IK}}{\epsilon _{LJ}}s\tilde s
		\end{equation}
		In particular
		\begin{equation}
			\left\{ \begin{array}{l}
		\displaystyle {\left| {{u^{IJ}}{Q_{IJ}}} \right|^2} = \frac{1}{2}\left( {{u^{IJ}}{u_{IJ}}} \right)s\tilde s\\[0.5em]
		{\left| Q \right|^2} =  - 2{{\hat u}_{IJ}}{{\hat u}^{IJ}} + s\tilde s
		\end{array} \right.
		\end{equation}
		
		(3)
		\begin{equation}
			R \cdot {Q_{IJ}} = s{u_{IJ}},\;\;\tilde R \cdot {Q_{IJ}} = \tilde s{u_{IJ}}.
		\end{equation}
		
		Positivity of the norms implies
		\begin{equation}
			s\tilde s \ge 2u_{IJ}u^{IJ} = 2{\hat u_{IJ}}{\hat u^{IJ}} + {u^2}.
		\end{equation}
When $s \tilde s = 2 u_{IJ} u^{IJ}$, we have $R$ and $\tilde R$ are parallel at such point, which in general we like to avoid.

		(4) Using Fierz identity, one can shows
		\begin{equation}
			\tilde sR + s\tilde R = 4{u_{IJ}}{Q^{IJ}} = 2uQ + 4{{\hat u}_{IJ}}{{\hat Q}^{IJ}},
		\end{equation}
		\begin{equation}
			\left\{ \begin{array}{l}
			{R_m}{{\tilde R}_n} - {R_n}{{\tilde R}_m} =  - 4{u_{IJ}}\chi _{mn}^{IJ}\;\;\; \Rightarrow \;\;\;\kappa  \wedge \tilde \kappa  =  - 4{u_{IJ}}{\chi ^{IJ}}\\[0.5em]
			\displaystyle {g_{mn}} =  - \frac{{2s\tilde s}}{{{{\left| {s\tilde R - \tilde sR} \right|}^2}}}\left[ {{R_m}{{\tilde R}_n} + {R_n}{{\tilde R}_m} - 4{{\left( {{Q_{IJ}}} \right)}_m}{{\left( {{Q^{IJ}}} \right)}_n}} \right]
\end{array} \right.,
		\end{equation}
where the last equation tells us that the metric is completely determined by the bilinears constructed from 2 solution. 

		\item \textbf{Contraction between the vectors and 2-forms}
		\begin{equation}
			\left\{ \begin{array}{l}
		{\iota _R}\left( {t\chi } \right) = s\left( {t\hat \tau } \right) - \left( {t\hat u} \right)\kappa \\[0.5em]
		{\iota _Q}({t\Theta}) = \left( {t\hat u} \right)\kappa  - s\left( {t\hat \tau } \right)\\[0.5em]
		{\iota _{\hat u\hat Q}}\left( {t\Theta } \right) = \left( {t\hat u} \right)\left( {u\kappa  + s\tau } \right)\\[0.5em]
		{\iota _R}( {{t^{IJ}}{{\tilde \Theta }_{IJ}}} ) - {\iota _{\tilde R}}\left( {{t^{IJ}}{\Theta _{IJ}}} \right) = 4{t^{IJ}}\left( {u{{\hat \tau }_{IJ}} - {{\hat u}_{IJ}}\tau } \right)
		\end{array} \right.
		\end{equation}
	where again $t_{IJ}$ is arbitrary triplet of functions.

	\end{itemize}

\section{Killing spinor equation}

	In this section we will discuss what constraints will be imposed on geometry of $M$ when there exists different number of solutions to the Killing spinor equation (\ref{Killing-equation}). We focus on situations where there are 1, 2, and 4 pairs of solutions to the equation.
	
	\vspace{20pt}
	Recall that the Killing spinor equation required by rigid limit of supergravity is
	\begin{equation}
		\delta {\psi _{mI}} = {\nabla _m}{\xi _I} - {\Gamma _m}{t_I}^J{\xi _J} - \frac{1}{2}{V^{pq}}{\Gamma _{mpq}}{\xi _I} - \frac{1}{2}{F_{mn}}{\Gamma ^n}{\xi _I} -{\left( {{A_m}} \right)_I}^J{\xi _J} = 0,
	\end{equation}
where $t_{IJ}$ is a triplet of scalars (or more precisely, a global section of the $ad(P_{SU(2)})$ where $P_{SU(2)}$ is an underlying principal $SU(2)_{\mathcal{R}}$-bundle, with gauge field ${(A_m)_I}^J$), $F$ is a closed 2-form, $V$ is a 2-form.

	The symplectic Majorana spinor $\xi_I$ is a section of the $SU(2)_{\mathcal{R}}$ twisted spin bundle of $M$. In general the $SU(2)_{\mathcal{R}}$-bundle $P$ is non-trivial. We define the gauge-covariant derivative on $t_{IJ}$
	\begin{equation}
		\nabla _m^A{t_I}^J \equiv {\nabla _m}{t_I}^J - {\left( {{A_m}} \right)_I}^K{t_K}^J + {t_I}^K{\left( {{A_m}} \right)_K}^J,
	\end{equation}
and curvature of $A$ as
	\begin{equation}
		{\left( {{W_{mn}}} \right)_I}^J \equiv {\nabla _m}{\left( {{A_n}} \right)_I}^J - {\nabla _n}{\left( {{A_m}} \right)_I}^J - \left[ {{{\left( {{A_m}} \right)}_I}^K{{\left( {{A_n}} \right)}_K}^J - {{\left( {{A_n}} \right)}_I}^K{{\left( {{A_m}} \right)}_K}^J} \right].
	\end{equation}
	
	Note that the Killing spinor equation is $SU(2)$ gauge covariant. It is also invariant under complex conjugation, provided that the auxiliary fields satisfies reality conditions: $F$ and $V$ are real,
	\begin{equation}
		\overline {{t_{IJ}}}  = {\epsilon ^{II'}}{\epsilon ^{JJ'}}{t_{I'J'}},
	\end{equation}
and similar for $A$.  The reality condition on $t_{IJ}$ and $A$ is just saying that they are linear combinations of Pauli matrices with imaginary coefficients.	

	Apart from the above obvious symmetries, the equation further enjoys a shifting symmetry and a Weyl symmetry.
	\begin{itemize}
		\item Shifting symmetry: The equation is invariant under the shifting transformation of auxiliary fields $V$ and $F$
		\begin{equation}
			\left\{ \begin{array}{l}
			V \to V + {\Omega ^ + }\\[0.5em]
			F \to F + 2{\Omega ^ + }
			\end{array} \right.,
		\end{equation}
	where $\Omega^+$ is any self-dual 2-form discussed in (\ref{self-duality}), following from the fact that
		\begin{equation}
			\Omega^+_{mn} \Gamma^{mn} \xi_I= 0.
		\end{equation}
	
		\item Weyl symmetry: after rescaling the metric $g \to e^{2\phi} g$, one can properly transform the auxiliary fields as well as the Killing spinor solution such that the Killing spinor equation is invariant. This can be seen by first rearranging the Killing spinor equation (\ref{Killing-equation}) into the form
		\begin{equation}
			{\nabla _m}{\xi _I} = {\Gamma _m}{{\tilde \xi }_I} + \frac{1}{2}{P_{mn}}{\Gamma ^n}{\xi _I},
			\label{reduced-Killing}
		\end{equation}
where
	\begin{equation}
		{{\tilde \xi }_I} \equiv \left( {{t_I}^J + \frac{1}{2}{V_{pq}}{\Gamma ^{pq}}\delta _I^J} \right){\xi _J}, \;\;{P_{mn}} \equiv {F_{mn}} - 2{V_{mn}}.
	\end{equation}
and we ignore the gauge field $A_{IJ}$ for simplicity.

	Focusing on (\ref{reduced-Killing}) alone as an equation for pair $(\xi, \tilde \xi)$ on any $d$-dimensional manifold, it is obvious that
	\begin{equation}
		{{\tilde \xi }_I} = \frac{1}{d}{\Gamma ^m}{\nabla _m}{\xi _I} - \frac{1}{2d}{P _{mn}}{\Gamma ^{mn}}{\xi _I}.
	\end{equation}
Substituting it back to (\ref{reduced-Killing}), one obtains the equation
	\begin{equation}
		\mathcal{D}(g){\xi _I} = \frac{1}{{2d}}{P_{pq}}{\Gamma _m}{\Gamma ^{pq}}{\xi _I} + \frac{1}{2}{P_{mn}}{\Gamma ^n}{\xi _I}
	\end{equation}
where the well-known differential operator $\mathcal{D}_g $ is defined as
	\begin{equation}
		\mathcal{D}(g) \equiv {\nabla _m} - \frac{1}{d}{\Gamma _m}{\Gamma ^n}{\nabla _n}.
	\end{equation}
and depends on the metric $g$. It's easy to show that\footnote{Under Weyl rescaling $g \to e^{2\phi}g$, the spin connection is shifted according to
	\begin{equation}
		{\nabla^g _m}\psi  \to \nabla^{{e^{2\phi}g} }_m \psi = {\nabla^g_m}\psi  + \frac{1}{2}\left( {{\nabla^g _n}\phi } \right){\Gamma _m}^n\psi .
	\end{equation}
To prove the Weyl transformation rule for $\mathcal{D}(g)$, one just need to plug the above formula into
	\begin{equation}
		D\left( {{e^{2\phi }}g} \right)\left( {{e^{\phi /2}}\psi } \right) = \nabla _m^{{e^{2\phi }}g}\left( {{e^{\phi /2}}\psi } \right) - \frac{1}{d}{\Gamma _m}{\Gamma ^n}\nabla _n^{{e^{2\phi }}g}\left( {{e^{\phi /2}}\psi } \right).
	\end{equation}
	}
	
	\begin{equation}
		{\mathcal{D}({{e^{2\phi }}g}}) {e^{ \phi /2}}= {e^{\phi /2}}{\mathcal{D}(g)}.
	\end{equation}
Hence, equation (\ref{reduced-Killing}) is invariant under rescaling
	\begin{equation}
		g \to e^{2\phi}g, \;\; P \to e^{\phi} P, \;\\; \xi \to e^{\phi/2} \xi.
	\end{equation}
	
	Now we return to the equation (\ref{Killing-equation}), and compute the transformation of auxiliary fields under Weyl rescaling. Suppose the scaling function $\phi$ is constant along vector field $R$:
	\begin{equation}
		{R^m}{\nabla _m}\phi  = 0,
	\end{equation}
then one can see that the Killing spinor equation (\ref{Killing-equation}) is invariant under rescaling
	\begin{equation}
		\boxed{
			g \to {e^{2\phi }}g,\;{t_{IJ}} \to {e^{ - \phi }}{t_{IJ}},\;V \to {e^\phi }V - \frac{e^\phi}{{2s}}\left( {\kappa  \wedge d\phi } \right),\;F \to {e^\phi }F - \frac{{{e^\phi }}}{s}\left( {\kappa  \wedge d\phi } \right)
		}\;,
	\end{equation}
provided we also rescale $\xi \to e^{\phi/2} \xi$. Note that the Weyl rescaling only affects the vertical part of $F$ and $V$. One can therefore use this rescaling symmetry with appropriate $\phi$ to make $F$ horizontal, namely
	\begin{equation}
		{\iota _R}F = 0.
	\end{equation}
However, unless explicitly stated, in most of the following discussions, we will keep the general $F$ without exploiting the Weyl symmetry.

	\end{itemize}

	\vspace{10pt}
	Let us comment on the reality condition defined earlier.
	
	(1) In 5 dimension Euclidean signature, the spinors belong to $2^2$ dimensional pseudoreal representation of ${\rm Spin}(5) \sim Sp(2)$, spinor $(\psi^*)_\alpha$ and $(C\psi)_\alpha \equiv C_{\alpha \beta} \psi^\beta$ transform in the same way. It is impossible to impose the usual Majorana condition, but one can impose the symplectic Majorana condition on spinors. In this sense, 4 complex (8 real) supercharges correspond to unbroken supersymmetry, namely $\mathcal{N} = 1$.
	
	The reality conditions introduced above are required by the supergravity that we started from, where one is interested in a real-valued action. However, it is fine to relax the reality condition on the Killing spinors and auxiliary fields, as long as one is only interested in a formally supersymmetric invariant theory. It makes perfect sense to consider complexified Killing spinor equation. In particular, the reality condition is not used in many of the following discussion, for instance, section 4.1 actually can be carried out without assuming the reality condition (except for the shifting symmetry of $\Omega^+$ which requires positivity following from reality condition). One only needs to work with $\mathbb{C}$-valued differential forms. Also, when we compare our 5d Killing spinor equation to the 4d equations appearing in \cite{Dumitrescu:2012ys}\cite{Dumitrescu:2012ly}, we drop the reality requirement. However, in this paper we mainly restrict ourselves to the real case, and reality condition does helps simplify certain discussions.
	
	(2) Solutions to equation (\ref{Killing-equation}) come in pairs. Suppose $\xi$ is a solution, corresponding to one supercharge $Q$, then its complex conjugate $\xi'$
	\begin{equation}
		{\xi '_1} = {\xi _2} = \overline {{\xi _1}} ,\;\;{\xi '_2} =  - {\xi _1},
	\end{equation}
automatically satisfies (\ref{Killing-equation}) corresponding to the supercharge $\overline Q$. The pair of solutions $\xi_I$ and $\xi'_I$ define the same scalar function $s$ and vector field $R$, but 2-forms $\Theta$ with different sign. 

	In view of such "pair-production" of solutions, we focus on finding different number of pairs of solutions to (\ref{Killing-equation}), and discuss them separately in the following subsections. When analyzing the case when $M$ admits 1 and 2 pairs of solutions, we will select one representative solution from each pair, say, $\xi$ and $\tilde \xi$, and study the relation between the bilinears that can be formed by these representing Killing spinors. Generically, the vector fields $R$ and $\tilde R$ from separate pairs should not be parallel everywhere on $M$.

	(3) One may worry about possible zeroes of Killing spinors. Similar to that in \cite{Dumitrescu:2012ys}, the Killing spinor equations are a first order homogeneous differential equation system, whose set of solutions span a complex vector space $\mathbb{C}^{k \le 4}$, with each solution completely specified by its value at a point $p\in M$. By the symplectic Majorana condition, $\xi_1(p) = 0$ implies $\xi_2 (p)= 0$, and hence $\xi_I (\forall x \in M) = 0$. Therefore, any non-trivial solution of the Killing spinor equation must be nowhere-vanishing, which ensures that the many bilinears defined (especially the almost contact structure) will be global.

	\vspace{10pt}
	In some sense, our Killing spinor equation is a generalization of the well-known Killing spinor equation
	\begin{equation}
		{\nabla _m}\psi  = \lambda {\Gamma _m}\psi,
	\end{equation}
The constant $\lambda$ can be real, pure-imaginary or zero, and the equation is accordingly called real, imaginary Killing spinor equation and covariantly constant spinor equation. If a manifold admits a Killing spinor, its Ricci curvature must take the form
	\begin{equation}
		Ric = 4\left( {n - 1} \right){\lambda ^2}g,
	\end{equation}
hence Einstein. For $\lambda$ pure imaginary, Baum gave a classification in \cite{Baum-1989}\cite{Baum-1989-2}. Prior to \cite{Bar}, manifolds with real Killing spinor are better known in low dimensions. For instance, 4-dimensional complete manifolds with real Killing spinor were shown to be isometric to the 4-sphere \cite{Friedrich}. In 5-dimension, simply-connected manifolds with real Killing spinors were shown to be round $S^5$ or Sasaki-Einstein manifolds, with solutions coming down from covariantly constant spinors on their Calabi-Yau cone. In \cite{Bar}, these results were generalized to higher dimensions: in dimension $n=4k+1$, only $S^{4n+1}$ and Sasaki-Einstein manifolds admits real Killing spinors, while in $n=4n+3 \ge 11$ dimension, only the round sphere, Sasaki-Einstein and 3-Sasakian manifolds admit real Killing spinors. 

	Our generalized Killing spinor equation has milder constraints on the geometry of manifold. We will see that the existence of one Killing spinor requires some soft geometry structure, one being an almost contact structure, similar to \cite{Closset:2012zr}. Of course, as the number of solutions increase, the geometry will be more constrained.

\subsection{Manifolds admitting 1 pair of supercharges}

	\subsubsection{General Result and ACMS structure}
	
	In this subsection we will analyze the case when there is one pair of solutions to the Killing spinor equation (\ref{Killing-equation}). We partially solve the auxiliary fields in terms of bilinears constructed, and rewrite the (\ref{Killing-equation}) into a simpler form. We will also briefly discuss 3 interesting cases with special auxiliary field configurations, which lead to geometrical restrictions of $M$ being locally foliated by special manifolds, or dimensional reduction to known 4d equations.
	
	\vspace{10pt}
	By differentiating the bilinears and using (\ref{Killing-equation}), one arrives at the following differential constraints on the quantities:
	
	\begin{itemize}
		\item Derivative on real positive function $s$
		\begin{equation}
			ds = - \iota_R F.
		\end{equation}
		
		\item Derivative on real vector field $R$
		\begin{equation}
			{\nabla _m}{{R}_n} =  2{\left( {t\Theta } \right)_{mn}} - \sqrt g {\epsilon ^{rpq}}_{nm}{{R}_r}{V_{pq}} + sF_{mn}.
		\end{equation}
		
		\item Derivative on the 2-form with any triplet $r_{IJ}$
		\begin{equation}
			\begin{array}{l}
			{\nabla _k}{\left( {{r ^{IJ}}{\Theta _{IJ}}} \right)_{mn}} = \left( {\nabla _k^A{r ^{IJ}}} \right){\left( {{\Theta _{IJ}}} \right)_{mn}}\\[0.5em]
			\;\;\;\;\;\;\;\;\;\;\;\;\;\;\;\;\;\;\;\;\;\;\;\;\;\;\;\; + {\rm{tr}}\left( {r t} \right)\left( {{g_{nk}}{R_m} - {g_{mk}}{R_n}} \right) - 2{r ^{JI}}{t_I}^K{\left( {*{\Theta _{JK}}} \right)_{kmn}}\\[0.5em]
			\;\;\;\;\;\;\;\;\;\;\;\;\;\;\;\;\;\;\;\;\;\;\;\;\;\;\;\; + 2\left[ {{{\left( {*V} \right)}_{nk}}^l{r ^{IJ}}{{\left( {{\Theta _{IJ}}} \right)}_{ml}} - {{\left( {*V} \right)}_{mk}}^l{{\left( {{r ^{IJ}}{\Theta _{IJ}}} \right)}_{nl}}} \right]\\[0.5em]
			\;\;\;\;\;\;\;\;\;\;\;\;\;\;\;\;\;\;\;\;\;\;\;\;\;\;\;\;  - {F_k}^p{r ^{IJ}}{\left( {*{\Theta _{IJ}}} \right)_{mnp}}
			\end{array}.
	\end{equation}
	
	\end{itemize}
	
	Let us comment on the above relations. The first equation implies $s = {\rm const}$ and can be normalized to $s=1$ when $F$ is horizontal. Recall that one can always use the Weyl symmetry of the equation to achieve this, although we keep the general situation. The second implies that $R$ is a Killing vector field:
	\begin{equation}
		\boxed{{\nabla _m}{R_n} + {\nabla _n}{R_m} = 0}.
	\end{equation}
The 3rd relation can be simplified as one puts in the solutions to $F$ and $V_H$.

	Using the 2nd and 3rd equation, one can solve (partially)  the auxiliary fields in terms of the bilinears (field $V$ is decomposed as $V = {V_H} + \kappa  \wedge \eta $) :
	\begin{equation}
		\boxed{
		\begin{array}{l}
		F = {\left( {2s} \right)^{ - 1}}d\kappa  + 2{s^{ - 1}}{\Omega ^ - } + 2{s^{ - 1}}{\Omega ^ + }\\[0.5em]
		{V_H} = - {s^{ - 1}}{(t\Theta)} + {s^{ - 1}}\Omega^- +  s^{-1} \Omega^+ \\[0.5em]
		\displaystyle {\eta ^m} = \frac{1}{{4{s^3}}}{\left( {{\Theta ^{IJ}}} \right)^{mn}}{\nabla ^k}{\left( {{\Theta _{IJ}}} \right)_{nk}} - \frac{3}{4}\left( {{\nabla ^m}{s^{ - 1}}} \right) - \frac{1}{{{s^2}}}{\left( {{A_n}} \right)_{IJ}}{\left( {{\Theta ^{IJ}}} \right)}^{nm}
		\end{array}
		}\;,
	\end{equation}
where $\Omega^\pm$ are self-dual ($+$) and anti-self-dual ($-$) 2-forms respectively, satisfying extra condition
	\begin{equation}
		{\mathcal{L}_R}\Omega^\pm  = 0.
	\end{equation}

	From previous discussions, we know that $\Omega^+$ corresponds to the arbitrary shifting symmetry of Killing spinor equation, so we may simply consider $\Omega^+ = 0$.

	$\Omega^-$ is in general non-zero. For instance, the well-known Killing spinor equation (\ref{Killing-Hosomichi}) corresponds to 
	\begin{equation}
		{\Omega ^ - } = - \frac{1}{4} d\kappa,
	\end{equation}
which is non-zero. Also, at the end of the paper we construct a supersymmetric theory for the $\mathcal{N} =1$ vector multiplet using the Killing spinor equation corresponding to
	\begin{equation}
		{\Omega ^ - } = \frac{1}{4}d\kappa .
	\end{equation}
However, to highlight some interesting underlying geometry related to (\ref{Killing-equation}), we will consider
	\begin{equation}
		\Omega^- = 0,
	\end{equation}
in this section unless explicitly stated. It is straight forward to generalize to non-zero $\Omega^-$, with sight modification to the following discussions.

	\vspace{10pt}
	
	Now that the auxiliary fields are partially solved, we can start simplifying the Killing spinor equation. As mentioned before, $t_{IJ}$ is a global section of associate rank-3 vector bundle of $P_{SU(2)}$, it may have zeroes. Below we will focus on 2 cases corresponding to $t \ne 0$ and $t = 0$ everywhere on $M$. 

	First let us consider the case when $t_{IJ} \ne 0$.
	
	(1) \underline{$t_{IJ} \ne 0$}
	
	Notice that the quantities $(g, s^{-1}R, s^{-1}\kappa, \varphi_t)$ actually form an {almost  contact metric structure} (abbreviated as ACMS). Using the ACMS, one can  further rewrite the Killing spinor equation:
	\begin{equation}
		\boxed{{{\hat \nabla }_m}{{\hat \xi }_I} - {({{\hat A}_m})_I}^J{{\hat \xi }_J} = 0},
		\label{Rewritten-Killing}
	\end{equation}
where we rescaled $\xi$
	\begin{equation}
		\hat \xi_I \equiv (\sqrt{s})^{-1} \xi_I,
	\end{equation}
	\begin{equation}
		\begin{array}{l}
		\displaystyle {({{\hat A}_m})_I}^J \equiv {\left( {{A_m}} \right)_I}^J + \frac{1}{s}{R_m}{t_I}^J + \frac{1}{{{\rm{tr}}\left( {{t^2}} \right)}}\left( {\nabla _m^A{t^{JK}}} \right){t_{KI}} + \eta \;{\rm{terms}}\\[0.5em]
		\displaystyle \;\;\;\;\;\;\;\;\;\;\;\; = \frac{1}{s}{R_m}{t_I}^J + \frac{1}{{{\rm{tr}}\left( {{t^2}} \right)}}\left( {{\nabla _m}{t^{JK}}} \right){t_{KI}} + \eta \;{\rm{terms}},
		\end{array}
	\end{equation}
and $\hat \nabla$ being the compatible spin connection introduced in the appendix \ref{Appendix-D}.
	\begin{equation}
		\begin{array}{l}
		\displaystyle {{\hat \nabla }_m}{\xi _I} = {\nabla _m}{\xi _I} + \frac{1}{{{\rm{tr}}\left( {{t^2}} \right)}}{({T_m})^J}_I{\xi _J} - \frac{1}{{2s}}{\nabla _m}{R_n}{\Gamma ^n}{\xi _I} + \frac{1}{2}\left( {{\nabla _m}\log s} \right){\xi _I}\\[0.5em]
		\displaystyle\;\;\;\;\;\;\;\;\;\;\;\;\; - \frac{1}{{{\rm{tr}}\left( {{t^2}} \right)}}{\eta _q}{\left( {t\Theta } \right)^q}_m{t_I}^J{\xi _J} + \frac{1}{2}{\left( {*{V^V}} \right)_{mpq}}{\Gamma ^{pq}}{\xi _I}
		\end{array}..
	\end{equation}	
	
	Notice that the new gauge connection is no longer $SU(2)$ connection, since the term
	\begin{equation}
		{\left( {{T_m}} \right)_{IJ}} \equiv \left( {\nabla _m^A{t_I}^K} \right){t_{KJ}},
	\end{equation}
might not be symmetric in $I,J$, but rather
	\begin{equation}
		{T_m^{IJ}} - {T_m^{JI}} = \frac{1}{2}{\epsilon ^{IJ}}{\nabla _m}{\rm{tr}}\left( {{t^2}} \right),
	\end{equation}
which corresponds to an new extra $U(1)$ gauge field. Fortunately this extra $U(1)$ part is in pure gauge,
	\begin{equation}
		\hat A_{U\left( 1 \right)}^{IJ}\sim{\epsilon ^{IJ}}\nabla\ln {\rm{tr}}\left( {{t^2}} \right),
	\end{equation}
and can be easily gauged away. Hence, let us choose a gauge
	\begin{equation}
		\nabla {\rm{tr}}\left( {{t^2}} \right) = 0.
	\end{equation}
	
	Before moving to the $t \equiv 0$ case, let us make a few remarks.
	
	(1) {The appearing of ACMS has already been hinted in literatures }. In \cite{Closset:2012zr}, supersymmetric theory is obtained on any 3d almost contact metric manifold. \cite{Kallen:2011ly} constructed twisted version of the super-Chern-Simons theory considered in \cite{Kapustin:2009aa} on any Seifert manifold $M_3$. Their twisted theory is defined with a choice of contact structure on $M_3$, with fermions replaced by differential forms. Note that the non-degenerate condition of a contact structure is crucial in defining the theory and the supersymmetry used for localization. Similar situations appear in \cite{Kallen:2012fk}\cite{Kallen:2012zr}, where the authors constructed twisted YM-CS theory on any 5d K-contact manifold $M$.

	(2) {There is an interesting configuration (among many similar ones)}. It corresponds to the case when
	\begin{equation}
		2V = F.
	\end{equation}
In such configuration,
	\begin{equation}
		d \kappa = - 4 t\Theta + 4 \kappa \wedge \eta \Rightarrow \kappa  \wedge d\kappa  \wedge d\kappa  \propto \kappa  \wedge \left( {t\Theta } \right) \wedge \left( {t\Theta } \right) \ne 0,
	\end{equation}
which implies $\kappa$ is a contact structure. To make things even simpler one can use the Weyl rescaling symmetry to make field $F$ as well as $V$ horizontal, and therefore $s = 1$:
	\begin{equation}
		F = \frac{1}{2}d\kappa  + 2{\Omega ^ - },\;\;V = \frac{1}{4}d\kappa  + {\Omega ^ - },
	\end{equation}
where $F$, $V$, $\Omega^-$ are now all closed anti-self-dual 2-forms. The Killing spinor equation can be rewritten as
	\begin{equation}
		{\nabla _m}{\xi _I} = {\Gamma _m}\left( {{t_I}^J + \frac{1}{4}{F^{pq}}{\Gamma _{pq}}\delta _I^J} \right){\xi _J},
	\end{equation}
which takes the familiar form
	\begin{equation}
		{\nabla _m}{\xi _I} = {\Gamma _m}{{\tilde \xi }_I},
	\end{equation}
with $\tilde \xi_I = ({t_I}^J + (1/4)F^{pq}\Gamma_{pq} \delta_I^J) \xi_J$. We will use this Killing spinor equation to construct a supersymmetric theory for the $\mathcal{N} = 1$ vector multiplet in section \ref{section-5}.

	There are many examples of contact manifolds. For instance, any non-trivial $U(1)$-bundle over a 4d Hodge manifold, with unit Reeb vector field $R$ pointing along the $U(1)$ fiber is a contact manifold. One should note that trivially fibered $S^1$-bundle, namely $M = S^1 \times N$ with Reeb vector field pointing along $S^1$ is not contact, because the non-degenerate condition cannot be satisfied. However, this type of manifold still serve as important examples admitting supersymmetry. Hence, we will have a brief discussion related to this type of manifold at the end of this section.

	\vspace{10pt}
	(2) \underline{$t_{IJ} \equiv 0$}.
	
	There is no natural ACMS arises in this case (although, if possible, one could choose by hand a nowhere-vanishing section of $ad(P_{SU(2)})$ to play the role of $t_{IJ}$, and similar calculations goes through. In this paper we do not consider this approach). The auxiliary fields $F$ and $V$ read
	\begin{equation}
		\left\{ \begin{array}{l}
		{F_{mn}} = {\left( {2s} \right)^{ - 1}}\left( {{\nabla _m}{R_n} - {\nabla _n}{R_m}} \right)\\[0.5em]
		{V_{mn}} = {R_m}{\eta _n} - {R_n}{\eta _m}
		\end{array} \right.,
	\end{equation}
and the Killing spinor equation reads
	\begin{equation}
		{\nabla _m}{{\hat \xi }_I} + \left[ { - \frac{1}{{4{s^2}}}\left( {{R_l}{\nabla _m}{R_n} - {R_n}{\nabla _m}{R_l}} \right) + \frac{1}{2}{{\left( {\iota_R *\eta} \right)}_{mnl}}} \right]{\Gamma ^{nl}}{{\hat \xi }_I} = {( {{A_m}\hat \xi } )_I}.
	\end{equation}
Similar to the previous discussion, we again have a new connection $\hat \nabla$ defined as
	\begin{equation}
		{{\hat \Gamma }^l}{_{mn}} = {\Gamma ^l}_{mn} + \frac{1}{{{s^2}}}\left( {{R^l}{\nabla _m}{R_n} - {R_n}{\nabla _m}{R^l}} \right) - 2{\left( {\iota_R * \eta} \right)^l}_{mn},
	\end{equation}
satisfying
	\begin{equation}
		{{\hat \nabla }_m}\left( {{s^{ - 1}}{R^n}} \right) = 0,
	\end{equation}
although there is no obvious geometrical interpretation for this connection.
	
	Again the Killing spinor equation can be rewritten as 
	\begin{equation}
		{{\hat \nabla }_m}{{\hat \xi }_I} = {\left( {{A_m}} \right)_I}^J{{\hat \xi }_J},
	\end{equation}
where $\hat \xi = \sqrt{s^{-1}} \xi$ has unit norm
	
	\vspace{10pt}
	To end this section, we discuss, in the following subsections, 3 special cases related to 5-manifolds of the form $M = S^1 \times M_4$, with the Reeb vector field $R$ pointing along $S^1$. As we will see there are 2 cases corresponding to two different types of auxiliary field configurations: $V$ horizontal, $F$ vertical and $V$, $F$ both vertical. The first configuration leads to geometric restrictions on the sub-manifold $M_4$, while the second corresponds to the dimensional-reduction of our 5d equation to 4d already discussed in the literatures.
	
	For such product form (or foliation) to appear, one first needs the horizontal distribution $TM_H$ to be integrable: the Frobenius integrability condition for $\kappa$ reads
	\begin{equation}
		d\kappa  \wedge \kappa  = 0\; {\rm , or\; equivalently}\; d\kappa  = \kappa  \wedge \lambda, \lambda \in \Omega^1_H(M).
	\end{equation}
Recall that $F \propto d \kappa$ ($\Omega^-$ is assumed to be $0$), one sees that the Frobenius integrability condition requires vertical $F$
	\begin{equation}
		F = \kappa \wedge (...).
	\end{equation}

	\subsubsection{Special Manifold foliation}

	\vspace{10pt}
	To proceed to the first class of special cases, let us define a local $SU(2)$ section of ``almost complex structure":
	\begin{equation}
		{J^a} \equiv \frac{i}{s}{\left( {{\sigma ^a}} \right)^I}_J{\Theta^J }_I,
	\end{equation}
satisfying
	\begin{equation}
		{J^a}{J^b} = {\epsilon ^{abc}}{J^c} - {\delta ^{ab}}I + {\delta ^{ab}} s^{-1}R \otimes s^{-1} \kappa .
	\end{equation}
It is immediate that when restricted on $TM_H$,
	\begin{equation}
		\boxed{
			{J^a}{J^b} = {\epsilon ^{abc}}{J^c} - {\delta ^{ab}}I
		}.
	\end{equation}
	
	Moreover, we have
	\begin{equation}
		\boxed{
			{\hat \nabla _k}{\left( {{J^a}} \right)_{mn}} = {( {{\hat A_k}} )^a}_b{\left( {{J^b}} \right)_{mn}}
			},
	\end{equation}
where
	\begin{equation}
		{( {{\hat A_m}} )^a}_b \equiv (-i)^2 {( {{\hat A_m}} )^I}_K{\left( {{\sigma ^a}} \right)^J}_I{\left( {{\sigma _b}} \right)^K}_J.
	\end{equation}
	
	Note that we can solve the new connection in terms of ``almost complex structures":
	\begin{equation}
		\boxed{{( {{{\hat A}_k}} )^a}_b = \frac{1}{4}{\left( {{J_b}} \right)^{mn}}{{\hat \nabla }_k}{\left( {{J^a}} \right)_{mn}}},
		\label{gauge-field}
	\end{equation}
which, depending on whether $t_{IJ} = 0$, provides constraints on $t_{IJ}$ or $A$.

	These equations closely resemble that of Quaternion-K\"ahler geometry, where one has on manifold $M$ a $SU(2)$ bundle of local almost complex structure $J^a$ satisfying
	\begin{equation}
		{J^a}{J^b} = {\epsilon ^{abc}}{J^c} - {\delta ^{ab}}I,
	\end{equation}
and is parallel with respect to the gauged connection
	\begin{equation}
		\nabla {J^a} = {A^a}_b{J^b},
	\end{equation}
with the Levi-civita connection $\nabla$ and a $SU(2)$ gauge connection $A$.
	
	However the situation here is slightly different. We do not have actually a manifold but rather a horizontal part of tangent bundle $TM_H$ of 5-fold $M$.
	
	Let us assume $V$ is horizontal:
	\begin{equation}
		\eta = 0.
	\end{equation}	

	The induced connection (for $t \ne 0$ case; $t=0$ case goes through similarly and yields the same conclusion) on $TM_H$ is
	\begin{equation}
		{\hat \nabla _X}Y = {\nabla _X}Y - g\left( { s^{-1} R,{\nabla _X}Y} \right) s^{-1}R - \frac{1}{{{\rm{tr}}\left( {{t^2}} \right)}}\left( {\nabla _X^A{t_I}^K} \right){t_{KJ}}{\Theta ^{IJ}}\left( Y \right), \; \forall X,Y \in TM_H
	\end{equation}
	
	Consider the special case where the sub-bundle $TM_H$ is integrable as the tangent bundle $T{M}_4$ of a co-dimension 1 sub-manifold ${M}_4$, then $\hat \nabla$ reduces to a connection on $M_4$. The first 2 terms of the connection combine to be the induced Levi-Civita connection $\nabla^{M_4}$ on $M_4$ ($s^{-1}R$ being the unit normal vector), while the third term add to it a torsion part:
	\begin{equation}
		{{\hat \Gamma }^n}{_{mk}} = {\Gamma ^n}_{mk} + {\gamma ^n}_{mk},
	\end{equation}
where
	\begin{equation}
		{\gamma ^n}_{mk} = - \frac{1}{{{\rm{tr}}\left( {{t^2}} \right)}}\left( {\nabla _m^A{t_I}^K} \right){t_{KJ}}{\left( {{\Theta ^{IJ}}} \right)^n}_k.
	\end{equation}
	
	Rewrite the Quaternion-K\"ahler-like equation as
	\begin{equation}
		\hat \nabla _k^{M_4} J_{mn}^a = \nabla _k^{M_4}J_{mn}^a - {\gamma ^l}_{km}J_{ln}^a - {\gamma ^l}_{kn}J_{ml}^a = {(\hat A_k)^a}_b J^b_{mn}.
	\end{equation}
	
	Now one can put back expression for both $\gamma$ and $J^a$, and sees that the torsion terms gives
	\begin{equation}
		{\gamma ^l}_{kn}J_{ml}^a - {\gamma ^l}_{km}J_{nl}^a = \frac{1}{{{\rm{tr}}\left( {{t^2}} \right)}}\left( {\nabla _k^A{t_I}^K} \right){t_K}^L{\left( {{\sigma ^a}} \right)_K}^J{\left( {{\sigma _b}} \right)^I}_J{\left( {{J^b}} \right)_{mn}} \equiv {(B_k)^a}_b (J^b)_{mn}.
	\end{equation}
This implies that the Quaternion-K\"ahler-like equation, restricted on a horizontal sub-manifold ${M_4}$, actually reduces to Quaternion-K\"ahler equation (with newer version of gauge field $\hat A + B$)
	\begin{equation}
		{\nabla ^{M_4}}{J^a} = {( {\hat A + B} )^a}_b{J^b} = \left ({(A_k)_I}^J+{R_k}{t_I}^J \right){\left( {{\sigma ^a}} \right)^I}_K{\left( {{\sigma _b}} \right)^K}_J{\left( {{J^b}} \right)_{mn}}.
	\end{equation}
Thus, we see that for generic auxiliary fields $t_{IJ}$ and $A_m$, provided that the horizontal distribution can be globally integrated to a sub-manifold $M_4$, $M_4$ is actually a {Quaternion-K\"ahler manifold}. Of course, there are special combinations of $t_{IJ}$ and $A$ such that $\hat A+B$ vanish. In such case, $M_4$ is a {HyperK\"ahler} manifold.

	With the integrability condition satisfied, we see that $M$ is now  locally foliated by Quaternion-K\"ahler (or HyperK\"ahler in special case) manifold. In particular, compact manifold $M$ could be a direct product 
	\begin{equation}
		\boxed{M = S^1 \times M_4,\; M_4 \;{\rm is \;Quaternion\; Kahler}}.
	\end{equation}
In view of the fact that there are only 2 compact smooth Quaternion-K\"ahler manifolds in $4d$, possible examples are $M = {S^1} \times \mathbb{C}{P^2},\;{S^1} \times {S^4}$, where the vector field $R$ is chosen to be the unit vector field along $S^1$, with gauge field $A$ turned on on $\mathbb{C}P^2$ and $S^4$. There are more examples when $M_4$ is allowed to be non-compact or orbifolds.
	
	\vspace{10pt}

	\subsubsection{Normal ACMS, Cosymplectic manifold and K\"ahler foliation}
	
	As mentioned above, there are 2 ways to define ACMS structure on $M$ using the data coming from Killing spinors: with the nowhere-vanishing auxiliary field $t_{IJ}$ or some other nowhere-vanishing section of $ad(P)$. In general the ACMS structure so defined does not have nice differential property. However, when some (rather strong) conditions are satisfied, the ACMS will behave nicer.
	
	\vspace{20pt}
	Let us focus on the case $t \ne 0$ and $(s^{-1}R, s^{-1}\kappa, \varphi_t)$ define ACMS on $M$.
	
	One obtains
	\begin{equation}
		{{\cal L}_R}t\Theta  = \frac{1}{2}\left( {\nabla _R^A{t^{IJ}}} \right)\left( {{\Theta _{IJ}}} \right) + s{\nabla ^p}\left( {\frac{1}{s}{R_m}} \right){\left( {t\Theta } \right)_{np}}d{x^m} \wedge d{x^n}.
	\end{equation}

	Setting
	\begin{equation}
		\nabla _R^At = 0,\;{\nabla _m}\left( {{s^{ - 1}}{R_n}} \right) = 0 \Leftrightarrow {\nabla _m}{R_n} \propto F_{mn} = 0,
	\end{equation} 	
one has $\mathcal{L}_R t\Theta = 0$ and hence ${\mathcal{L}_{{s^{ - 1}}R}}{\varphi _t} = 0$.

	If, in additional to the above, one further imposes $V$ to be horizontal and $\nabla^A t = 0$, then it is easy to see that the ACMS satisfies
	\begin{equation}
		\nabla {\varphi _t} = 0,
	\end{equation}
and hence it is cosymplectic. In this case, the Levi-civita connection $\nabla$ on $M$ respects the ACMS, the restriction of $\nabla$ on the horizontal distribution is automatically a connection on $TM_H$.
	
	Note that $\nabla R = 0$ implies that the horizontal distribution is locally integrable. Therefore, restricted on the integral sub-manifold, $\nabla$ is the induced Levi-civita connection, $\varphi_t$ is an almost complex structure which can be shown to have vanishing Nijenhuis tensor and hence actually a complex structure. It is parallel with respect to induced Levi-civita connection, hence is K\"ahler.
	
	In summary, we see that
	\begin{equation}
		{\nabla ^A}{t^{IJ}} = 0,\;F = 0,\;V  = V_H = - t\Theta,
	\end{equation}
implies a cosymplectic ACMS (namely $\nabla \varphi_t = 0$), and $M$ is locally foliated by 4d K\"ahler manifold, with the K\"ahler structure provided by $\varphi_t$.

	Recall that we had conclusion that $M$ is locally foliated by Quaternion-K\"ahler manifold in the previous subsection, for configuration $F_H = 0$, $V=V_H$. Suppose $M= M_4\times S^1$ with a Reeb vector field $R$ from a Killing spinor pointing along $S^1$, then we see that $M_4$ must be Quaternion-K\"ahler as well as K\"ahler. If $M_4$ is a smooth compact manifold, then this leaves only one possibility:
	\begin{equation}
		M = \mathbb{CP}^2 \times {S^1}.
	\end{equation}
Of course, for more general Reeb vector field pointing along other directions, one could have other possibilities of $M_4$.

	\vspace{10pt}

	\subsubsection{Reducing to 4d}
	
	Finally let us point out the reduction of (\ref{Killing-equation}) to 4d already discussed in literatures\cite{Dumitrescu:2012ys}\cite{Dumitrescu:2012ly}. Consider  $M = M_4 \times S^1$ with spinor $\xi_I$ and auxiliary fields independent on the $S^1$ coordinate. The 4d part of the Killing spinor equation reads
	\begin{equation}
		{\nabla _\mu }{\xi _I} = {t_I}^J{\gamma _\mu }{\xi _J} + \frac{1}{2}{F_{\mu 5}}{\gamma ^5}{\xi _I} + \frac{1}{2}{V^{\nu 5}}{\gamma _{\mu \nu 5}}{\xi _I} + \frac{1}{2}{V^{\lambda \rho }}{\gamma _{\mu \lambda \rho }}{\xi _I} + \frac{1}{2}{F_{\mu \nu }}{\gamma ^\nu }{\xi _I} + {\left( {{A_\mu }} \right)_I}^J{\xi _J},
	\end{equation}
and the $S^1$ part serves as direct constraints on auxiliary fields
	\begin{equation}
		{\partial _5}{\xi _I} = {t_I}^J{\xi _J} + \frac{1}{2}{F_{5\mu }}{\gamma ^\mu }{\xi _I} + \frac{1}{2}{V^{\mu \nu }}{\gamma _{\mu \nu }}{\gamma _5}{\xi _I} + {\left( {{A_5}} \right)_I}^J{\xi _J} = 0.
	\end{equation}
	
	There are now 2 different ways to reduce the equation, each gives rise to the Killing equation discussed in \cite{Dumitrescu:2012ys}\cite{Dumitrescu:2012ly}. The involved vertical condition $V_H = F_H = 0$ and requirement $t = 0$ or $t_{IJ} \propto \epsilon_{IJ}$ indeed imply the Frobenius Integrability condition
	\begin{equation}
		d\kappa  \wedge \kappa  = 0,
	\end{equation}
which is necessary for $M$ to be a product.
	
	\underline{I. Reduction to \cite{Dumitrescu:2012ys}}
	
	Setting $t = A = F_{\mu \nu} = V_{\mu \nu} = 0$, namely $F$ and $V$ are both vertical 2-forms, the equation simplifies to
	\begin{equation}
		\left\{ \begin{array}{l}
		\displaystyle {\nabla _\mu }{\xi _I} = \frac{1}{2}{F_{\mu 5}}{\gamma ^5}{\xi _I} + \frac{1}{2}{V^{\nu 5}}{\gamma _{\mu \nu 5}}{\xi _I}\\[0.5em]
		{\partial _5}{\xi _I} = {F_{5\mu }}{\gamma ^\mu }{\xi _I} = 0
		\end{array} \right.,
	\end{equation}
or written in terms of Weyl components ${\xi _I} = ( {{\zeta _I},{{\tilde \zeta }_I}} )$, 
	\begin{equation}
		\left\{ \begin{array}{l}
		\displaystyle {\nabla _\mu }{\zeta _I} = \frac{1}{2}{F_{\mu 5}}{\zeta _I} + \frac{1}{2}{V^{\nu 5}}{\sigma _{\mu \nu }}{\zeta _I}\\[0.5em]
		\displaystyle{\nabla _\mu }{{\tilde \zeta }_I} =  - \frac{1}{2}{F_{\mu 5}}{\tilde \zeta _I} - \frac{1}{2}{V^{\nu 5}}{\tilde \sigma _{\mu \nu }}{\tilde \zeta _I}
		\end{array} \right.,
	\end{equation}
with constraint on $F_{\mu 5}$
	\begin{equation}
		{F_{5\mu }}{{\tilde \sigma }^\mu }\zeta_I  = 0,\;{F_{5\mu }}{\sigma ^\mu }\tilde \zeta_I  = 0.
	\end{equation}
	
	Suppose we relax the reality condition on $\xi$ and also $F$ and $V$, and define new complex auxiliary vector fields $A$ and $V$
	\begin{equation}
		\left\{ \begin{array}{l}
		2i{A_\mu } \equiv {F_{\mu 5}} - {V_{\mu 5}} = \partial_\mu a_5 - V_{\mu 5}\\[0.5em]
		- 2i{V_\mu } \equiv {V_{\mu 5}}
		\end{array} \right.	,
	\end{equation}
then the above equation takes a familiar form
	\begin{equation}
		\left\{ \begin{array}{l}
{\nabla _\mu }{\zeta _I} =  - i\left( {{V_\mu } - {A_\mu }} \right){\zeta _I} - i{V^\nu }{\sigma _{\mu \nu }}{\zeta _I}\\[0.5em]
		{\nabla _\mu }{{\tilde \zeta }_I} = i\left( {{V_\mu } - {A_\mu }} \right){{\tilde \zeta }_I} + i{V^\nu }{\tilde \sigma _{\mu \nu }}{{\tilde \zeta }_I}
\end{array} \right.,
	\end{equation}
which is just the Killing equations discussed in \cite{Dumitrescu:2012ys} for 2 separate pairs of chiral spinors  $(\zeta_1, \tilde \zeta_1)$ and $(\zeta_2, \tilde \zeta_2)$. $V_{\mu 5}$ has to satisfy conservation condition $\nabla_\mu V^{\mu 5} = 0$, and $F_{\mu 5}$ is holomorphic w.r.t $J^I_{\mu \nu}$ and $\tilde J^I_{\mu \nu}$ if any of them is non-zero. The conservation condition on $V_{\mu 5}$ is equivalent to $d^*$-closed condition on vertical 2-form $V$
	\begin{equation}
		{\nabla _\mu }{V^{\mu 5}} = 0 \Leftrightarrow {\nabla ^m}{V_{mn}} = 0 \Leftrightarrow d* V = 0.
	\end{equation}

	Now that we choose not to impose reality condition on auxiliary fields, it is also fine for $\xi_I$ to be non-sympletic-Majorana, hence $\xi_1$ and $\xi_2$ are now unrelated complex spinors, and one of the two can vanish. This then leads to different numbers of Killing spinor solutions in 4d, ranging from 1 to 4. In \cite{Dumitrescu:2012ys}, the cases when $M_4$ admits 1, 2 and 4 supercharges are discussed in detail. Here we list a few points and discuss their 5d interpretation. More results can be obtained similarly.
	
	\underline{(1) 2 supercharges of the form $(\zeta, 0)$ and $(\eta, 0)$}: then assuming $M_4$ is compact,  $M_4$ has to be a Hyperhermitian manifold up to global conformal transformation. Moreover, the auxiliary fields satisfy
	\begin{itemize}
		\item a) ${V_\mu } - {A_\mu }$ is closed 1-form.
		\item b) $\partial_\mu V_\nu - \partial_\nu V_\mu$ is anti-self-dual 2-form.
	\end{itemize}
	
	Condition a) is obviously satisfied by definition: $V_{\mu} - A_{\mu}\sim \partial_\mu{a_5}$ is obviously closed. The condition b) reads in 5d point of view
	\begin{equation}
		{\iota _R}dV =  - * dV,
	\end{equation}
	
	\underline{(2) 2 supercharges of the form $(\zeta, 0)$ and $(0, \tilde \zeta)$}: there are 2 commuting Killing vector on $M_4$, and hence $M_4$ is locally $T^2$-fibration over Riemann surface $\Sigma$. The auxiliary fields $V_{\mu 5}$ and $F_{\mu 5}$ are given in terms of $J_{\mu \nu}$ and $\tilde J_{\mu \nu}$.

	\vspace{10pt}
	\underline{II. Reduction to \cite{Dumitrescu:2012ly}}
	
	Setting $A = F_{\mu \nu} = V_{\mu \nu} = 0$, $\displaystyle \frac{1}{2}{F_{\mu 5}} = \frac{1}{2}{V_{\mu 5}} = \frac{i}{3}{b_\mu }$, $t = (i/6) M I_{2\times 2}$, one similarly obtains
	\begin{equation}
		\left\{ \begin{array}{l}
		\displaystyle {\nabla _\mu }{\zeta _I} = \frac{i}{6}M{{\tilde \sigma }_\mu }{{\tilde \zeta }_I} + \frac{i}{3}{b_\mu }{\zeta _I} + \frac{i}{3}{b^\nu }{\sigma _{\mu \nu }}{\zeta _I}\\[0.8em]
		\displaystyle{\nabla _\mu }{{\tilde \zeta }_I} = \frac{i}{6}M{\sigma _\mu }{\zeta _I} - \frac{i}{3}{b_\mu }{{\tilde \zeta }_I} - \frac{i}{3}{b^\nu }{\tilde \sigma _{\mu \nu }}{{\tilde \zeta }_I}
		\end{array} \right.,
	\end{equation}
which is the Killing spinor equation for 2 pairs of spinor $(\zeta_1, \tilde \zeta_1)$ and $(\zeta_2, \tilde \zeta_2)$ discussed in \cite{Dumitrescu:2012ly} for but with condition $M = \tilde M$. 

	Again, $\xi_I$ are no longer symplectic Majorana, and solution of the 5d Killing spinor equation leads to different number of solutions to 4d Killing spinor equation. Let us list a few examples from the detail discussion in \cite{Dumitrescu:2012ly}. Interested reader can refer to their paper for more results.
	
	(1) \underline{1 supercharge of the form $(\zeta, \tilde \zeta)$}: Any manifold $(M_4, g)$ with a nowhere-vanishing complex Killing vector field $K$ which squares to zero and commutes with its complex conjugate
	\begin{equation}
		K_\mu K^\mu =0, \; [K, \bar K] = 0,
	\end{equation}
admits solution of the form $(\zeta, \tilde \zeta)$ to the 4d Killing spinor equation. $K$ and the metric can be used to build up a Hermitian structure on $M_4$. 
	
	(2) \underline{2 supercharges of the form $(\zeta_1, 0)$ and $(\zeta_2,0)$}: $M_4$ is anti-self-dual with $V_{5\mu}$ and $F_{5\mu}$ closed 1-forms, and hence in 5d point of view, they are closed vertical 2-forms. Moreover, the form of solution requires $\tilde M = 0$, and according to our reduction, $M = \tilde M =0$. If $F = V$ are exact, then $M_4$ is locally conformal to a Calabi-Yau 2-fold. Otherwise, $M_4$ is locally conformal to $\mathbb{H}^3 \times \mathbb{R}$.
	
	(3) \underline{2 supercharges of the form $(\zeta_1, 0 )$ and $(0, \tilde \zeta_2)$}: One must have $M = \tilde M = 0$. This is equivalent to $M_4$ having solution $(\zeta_1, \tilde \zeta_2)$ with $M = \tilde M = 0$.

\subsection{Manifolds admitting 2 pairs of supercharges}

	In this section we consider the case when 2 pairs of solutions to the (\ref{Killing-equation}) exist. We see that when certain assumptions on vectors $Q_{IJ}$ are made, and if the Killing vector fields form closed algebra, the geometry of $M$ will be heavily constrained. And in particular, all the resulting geometries admit contact metric structures.
	
	\vspace{10pt}
	The spinors $\xi$ and $\tilde \xi$ satisfy equations:
	\begin{equation}
		\begin{array}{l}
		\displaystyle {\nabla _m}{\xi _I} = {t_I}^J{\Gamma _m}{\xi _J} + \frac{1}{2}{V^{pq}}{\Gamma _{mpq}}{\xi _I} + \frac{1}{2}{F_{mp}}{\Gamma ^n}{\xi _I} + {\left( {{A_m}} \right)_I}^J{\xi _J}\\[0.5em]
		\displaystyle{\nabla _m}{{\tilde \xi }_I} = {t_I}^J{\Gamma _m}{{\tilde \xi }_J} + \frac{1}{2}{V^{pq}}{\Gamma _{mpq}}{{\tilde \xi }_I} + \frac{1}{2}{F_{mn}}{\Gamma ^n}{{\tilde \xi }_I} + {\left( {{A_m}} \right)_I}^J{{\tilde \xi }_J}
\end{array}.
	\end{equation}

	Similar to the previous section, we have
	\begin{itemize}
		\item Derivative on $u_{IJ}$

		(1)
		\begin{equation}
			{u^{IJ}}d{u_{IJ}} = {{\hat u}^{IJ}}d{{\hat u}_{IJ}} + \frac{1}{2}udu =  - 2{t^{IJ}}{\left( {\hat u\hat \tau } \right)_{IJ}} - {\iota _{(uQ)}}F.
		\end{equation}
		
		(2)
		\begin{equation}
			du =  - {\iota _Q}F.
		\end{equation}
		
		\item Derivative on $Q_{IJ}$
		\begin{equation}
			\boxed{
				{\nabla _m}{Q_n} + {\nabla _n}{Q_m} = 0
				}.,
		\end{equation}
	namely, $Q$ is a Killing vector.

	The derivative on $u_{IJ}$ implies relation
	\begin{equation}
		2{u_{IJ}}{u^{IJ}} = s\tilde s + C,
	\end{equation}
where the function $C$ is invariant along $R$ and $\tilde R$. When $t_{IJ} = 0$, $C$ reduces to constant. Notice that when $C = 0$,
	\begin{equation}
		s\tilde R = \tilde sR,
	\end{equation}
and when $C = -s\tilde s$
	\begin{equation}
		\tilde s R = - s \tilde R,
	\end{equation}
which are degenerate cases that we do not consider in the following.
		\item Commutator between $R$ and $\tilde R$
		\begin{equation}
			K\equiv {[ {R,\tilde R} ]^m} = 8\left( {t\hat u} \right){Q^m} - 8u{( {t\hat Q} )^m} - 4{\left( {{\iota _R}{\iota _{\tilde R}}*V} \right)^m} + {\left( {\tilde s{\iota _R}F - s{\iota _{\tilde R}}F} \right)^m}.
	\end{equation}
	
	\end{itemize}

	\vspace{20pt}
	Recall that we now have several Killing vector fields, $R$, $\tilde R$, $K$ and $Q$. If some of them form closed Lie algebra, the geometry of $M$ will be heavily constrained. In the rest of this section, we discuss several simplest possibilities where they form 2 or 3 dimensional Lie algebras.

	\underline{1. $R$ and $\tilde R$ form 2-dimensional algebra}
	
	There exist only two 2-dimensional Lie algebras up to isomorphisms. One is the abelian algebra, the other is a unique non-abeilian algebra.
	
	When $R$ and $\tilde R$ commute, namely $K= 0$, one obtains the abelian algebra. If the orbits of $R$ and $\tilde R$ are closed, then $M$ is acted freely by $T^2$, and therefore $M$ is a $T^2$-fibration.
	
	The non-abelian algebra corresponds to $K \ne 0$. Assume $K$ is a linear combination of $R$ and $\tilde R$, then
	\begin{equation}
		[R, \tilde R] = aR + b \tilde R.
	\end{equation}
Contracting with $R$ and $\tilde R$ it gives
	\begin{equation}
		\left\{ \begin{array}{l}
		a{s^2} + b\left( {s\tilde s + 2C} \right) = s{\iota _{\tilde R}}{\iota _R}F\\[0.5em]
		a\left( {s\tilde s + 2C} \right) + b{{\tilde s}^2} = \tilde s{\iota _{\tilde R}}{\iota _R}F
		\end{array} \right. .
	\end{equation}
The determinant of the system is
	\begin{equation}
		\det  = {s^2}{{\tilde s}^2} - {\left( {s\tilde s + 2C} \right)^2} =  - 4C\left( {C + s\tilde s} \right).
	\end{equation}
Notice that away from the degenerate cases when $C = 0$ and $C = -s\tilde s$, the determinate is non-zero. Therefore, when $\iota_R \iota_{\tilde R} F \ne 0$, the system allows solution $(a,b)$
	\begin{equation}
		\left\{ \begin{array}{l}
		\displaystyle b = \frac{{s{\iota _{\tilde R}}{\iota _R}F}}{{2\left( {s\tilde s + C} \right)}}\\[0.8em]
		\displaystyle a = \frac{{\tilde s{\iota _{\tilde R}}{\iota _R}F}}{{2\left( {s\tilde s + C} \right)}}
		\end{array} \right..
	\end{equation}
	
	Notice however that $R$, $\tilde R$ and their commutator are all Killing vectors, therefore the coefficients $a$ and $b$ must be constant. This implies
	\begin{equation}
		\frac{s}{{\tilde s}} = {\rm const},
	\end{equation}
and further 
	\begin{equation}
		{\mathcal{L}_R}\tilde s = {\mathcal{L}_{\tilde R}}s = 0 \Rightarrow {\iota _R}{\iota _{\tilde R}}F = 0,
	\end{equation}
hence
	\begin{equation}
		a = b = 0.
	\end{equation}
	
	To summarize, if $R$ and $\tilde R$ form 2-dimensional algebra, it can only be trivial abelian algebra.
	
	What remains is the Killing vector $Q$. Assume $Q$ and the commuting $R$ and $\tilde R$ form 3 dimensional algebra:
	\begin{equation}
		\left\{ \begin{array}{l}
		[ {R,\tilde R} ] = 0\\[0.5em]
		\left[ {Q,R} \right] = aR + b\tilde R + mQ\\[0.5em]
		[ {Q,\tilde R} ] = cR + d\tilde R + nQ
		\end{array} \right..
	\end{equation}
	
	Let us make a Weyl rescaling to set $\iota_R F = 0$. Then it automatically implies $\iota_R \iota_Q F = \iota_{\tilde R} \iota_Q F = 0$ by previous arguments. Therefore,
	\begin{equation}
		\left\{ \begin{array}{l}
		{\mathcal{L}_R}\left( {u\tilde s} \right) = {\mathcal{L}_R}( {\tilde R \cdot Q} ) = \tilde R \cdot \left[ {R,Q} \right] = 0\\[0.5em]
		{\mathcal{L}_{\tilde R}}\left( {us} \right) = {\mathcal{L}_{\tilde R}}\left( {R \cdot Q} \right) = R \cdot [\tilde R,Q] = 0
		\end{array} \right..
	\end{equation}
It is immediate to see that the determinant of the above linear system is
	\begin{equation}
		\det  \propto | {s\tilde R - \tilde sR} |^2{\left| Q \right|^2},
	\end{equation}
and hence non-trivial solution requires $Q=0$ or $\tilde s R = s \tilde R$, which we do not consider. Therefore, one has $Q, R, \tilde R$ forming abelian algebra, and $M$ is a $T^3$-fibration over Riemann surface $\Sigma$. Up to an overall rescaling factor, the metric can be written as
	\begin{equation}
		d{s^2} = {h_{\alpha \beta }}d{x^\alpha }d{x^\beta } + \sum\limits_{i = 1}^3 {{{\left( {d{\theta _i} + {\alpha _i}(x)} \right)}^2}} ,
	\end{equation}
where $\theta_i$ are the periodic coordinates along $R$, $\tilde R$ and $Q$ provided their orbits are closed, and $\alpha_i$ are 1-forms that determine the fibration.
	
	\vspace{20pt}
	\underline{2. $R$, $\tilde R$ and $[R, \tilde R]$ form 3-dimensional algebra}
	
	Assume that the algebra takes the form
	\begin{equation}
		\left\{ \begin{array}{l}
		[ {R,\tilde R} ] = K\\[0.5em]
		\left[ {R,K} \right] = aR + b\tilde R + mK\\[0.5em]
		[ {\tilde R,K} ] = cR + d\tilde R + nK
		\end{array} \right..
	\end{equation}

	In general, $\iota_R \iota_{\tilde R} F$ does not vanish. However, we can make a Weyl rescaling to make, for instance, $\iota_R F =0$, and in particular, $s$ is constant and $\iota_R \iota_{\tilde R} F= 0$. This implies
	\begin{equation}
		R \cdot K = \tilde R \cdot K = 0.
	\end{equation}	
	
	It is then easy to solve the coefficients in the above linear relation:
	\begin{equation}
		\left\{ \begin{array}{l}
		\displaystyle a =  - \frac{1}{{4C}}{\left| K \right|^2}\frac{{s\tilde s + 2C}}{{s\tilde s + C}}\\[0.8em]
		\displaystyle b = \frac{1}{{4C}}{\left| K \right|^2}\frac{{{s^2}}}{{s\tilde s + C}}
		\end{array} \right.,\;\;\left\{ \begin{array}{l}
		\displaystyle c =  - \frac{1}{{4C}}{\left| K \right|^2}\frac{{{s^2}}}{{s\tilde s + C}}\\[0.8em]
		\displaystyle d = \frac{1}{{4C}}{\left| K \right|^2}\frac{{s\tilde s + 2C}}{{s\tilde s + C}}
		\end{array} \right..
	\end{equation}
	
	The fact that all coefficients must be constants implies
	\begin{equation}
		\frac{s}{{\tilde s}} = {\rm{const,}}\;\;\frac{{{s^2}}}{{s\tilde s + 2C}} = {\rm{const}},
	\end{equation}
and therefore both $\tilde s$ and $C$ are constant as well.

	It is then straight forward to renormalize and linearly recombine the vectors to form a standard $\mathfrak{su(2)}$ algebra. Therefore topologically $M$ is a $SU(2)$-fibration over a Riemann surface $\Sigma$; however, there is no non-trivial $SU(2)$ bundle over a Riemann surface from the fact that the 3-skeleton of the classifying space $BSU(2)$ is a point), hence topologically $M = S^3 \times \Sigma$. Up to an overall scaling factor which was used to bring $s$ to 1, the metric takes the form
	\begin{equation}
		ds_M^2 = ds_\Sigma ^2 + ds_{{S^3}}^2 = {h_{\alpha \beta }}(x)d{x^\alpha }d{x^\beta } + \sum\limits_{a = 1}^3 {{e^a}{e^a}} ,
	\end{equation}
where $e^a = \kappa, \tilde \kappa, \gamma$ are $SU(2)$ invariant 1-forms on $SU(2)$. Note that $\iota_R F = \iota_{\tilde R} F = 0$ implies $F$ is a form on $\Sigma$:
	\begin{equation}
		F = \frac{1}{2}{F_{\alpha \beta }}\left( x \right)d{x^\alpha } \wedge d{x^\beta }.
	\end{equation}
	
	Recall that there is one more Killing vector field $Q$. The metric has isometry subgroup $SU(2)_L \times SU(2)_R$, which comes from the isometry of $S^3$. If $Q \notin \mathfrak{su}(2)_L \times \mathfrak{su}(2)_R$, then $Q$ must generate continuous isometry in $\Sigma$, which implies $\Sigma = T^2$ or $S^2$ if $M$ is compact. In this case, by requiring $Q$ commutes and being orthogonal to $R$, $\tilde R$ and $K$, one can derive new constraints on the auxiliary fields. For instance,
	\begin{equation}
		R \cdot Q = 0 \Leftrightarrow u = 0 \Leftrightarrow {\iota _Q}F = 0
	\end{equation}
which, combining with the fact that $F$ is a 2-form on $\Sigma$, implies actually $F = 0$.

\subsection{Manifolds admitting 8 supercharges}
	
	In this section, we discuss the optimal case where the Killing spinor equation has full 4 complex dimensional space of solutions. This is done by taking the commutator of the $\nabla$, applying Killing spinor equation and matching the Gamma matrix structure on both sides. We will see that there are 3 cases corresponding to the survival of only one of the 3 auxiliary fields $(t, V, F)$, with the other two vanishes identically. Here we list main results that we will discuss in detail:
	
	\begin{itemize}
		\item $V \ne 0$: $M$ is positively curved, with product structure $T^k \times G$ where $G$ is a compact Lie group. The non-trivial example is then $T^2 \times SU(2)$ with standard bi-invariant metric.
		
		\item $F \ne 0$: $M$ is locally of the form $M_3 \times \mathbb{H}^2$, where $M_3$ is a 3 dimensional flat manifold.
		
		\item $t \ne 0$: $M$ is locally a space of constant curvature with positive scalar curvature, hence $M$ is locally isometric to $S^5$.
		
		\item $t=V=F=0$: $M$ has zero curvature, hence is locally isometric to $\mathbb{R}^5$.
		
	\end{itemize}

	\vspace{20pt}
	By explicitly writing down the commutator $[\nabla_m, \nabla_n] \xi_I$ using Killing spinor equation, one would obtain 2 immediate results:
	\begin{itemize}
	\item Terms independent of $\Gamma$.
		\begin{equation}
		\boxed{{\left( {{W_I}^J} \right)_{mn}} \equiv {\nabla _m}{\left( {{A_n}} \right)_I}^J - {\nabla _n}{\left( {{A_m}} \right)_I}^J + {\left( {{A_n}} \right)_I}^K{\left( {{A_m}} \right)_K}^J - {\left( {{A_m}} \right)_I}^K{\left( {{A_n}} \right)_K}^J = 0}.
		\end{equation}
For simply-connected 5-manifolds, flat connections must be gauge equivalent to trivial connections.
		
	\item Terms linear in $\Gamma$.
		\begin{equation}
		\begin{array}{l}
		0 = \left( {{\nabla _m}{t_I}^J} \right){\Gamma _n} - \left[ {\left( {{A_m}t} \right)_I}^J - {\left( {t{A_m}} \right)_I}^J \right] \Gamma_n\\[0.5em]
		\displaystyle \;\;\;\;\;\; + \frac{1}{2} ({\nabla _m}{F_{np}}){\Gamma ^p}{\delta _I}^J - {F_n}^p{\left( {*V} \right)_{mp}}{\;^q}{\Gamma _q}{\delta _I}^J - 2{t_I}^J{\left( {*V} \right)_{mn}}^l{\Gamma _l}\\[0.5em]
		\;\;\;\;\;\; - \left( {m \leftrightarrow n} \right)
		\end{array}.
		\label{linear-in-gamma}
		\end{equation}
	\end{itemize}
The solutions to the equation are:

	\textbf{Case 1}
		\begin{equation}
			\left\{ \begin{array}{l}
			{t_{IJ}} = 0\\
			F = 0
			\end{array} \right.
		\end{equation}
		
	\textbf{Case 2}
		\begin{equation}
			V=0
		\end{equation}

	Now we study 2 cases separately.
	\vspace{10pt}
	
	\underline{\textbf{Case 1}: Only $V \ne 0$}.
	
	The solution $t^{IJ} = 0$ and $F = 0$ implies (\ref{linear-in-gamma}) vanishes identically, no further condition on $V$ is required.
	
	Combining with previous section, we know that
	\begin{equation}
		ds = 0,
	\end{equation}
and we conveniently set $s = 1$.
	
	By identifying the terms quadratic in $\Gamma$ matrices, one sees that the 
	
	\begin{itemize}
		\item The curvature tensor satisfies a flat condition:
	\begin{equation}
		{\hat R_{mnkl}} (\hat \nabla) = 0,
	\end{equation}
where $\hat R$ is the curvature tensor of a metric connection with anti-symmetric torsion
	\begin{equation}
		{\hat \nabla _m}{X^n} = {\nabla _m}{X^n} + 2{\left( {*V} \right)^n}_{mk}{X^k}.
	\end{equation}
with $\nabla$ the Levi-civita connection of $g$. This result is most easily understood by looking at the Killing spinor equation, where $V$ can be absorbed into the metric connection as a totally anti-symmetric torsion.

		\item The Ricci curvature
		\begin{equation}
			Ri{c_{mn}} =  4{\left( {*V} \right)^{pq}}_m{\left( {*V} \right)_{pqn}}.
		\end{equation}

		\item Scalar curvature
		\begin{equation}
			\mathcal{R} =  +4 {\left( {*V} \right)^{kmn}}{\left( {*V} \right)_{kmn}} \ge 0,
		\end{equation}
which indicates the manifold must have {positive} curvature. Moreover, compact manifolds admitting metric connection with anti-symmetric torsion are known to be products of $T^k \times G$ where $G$ is a compact group. This leaves us only a few possibilities, the non-trivial one being
	\begin{equation}
		M = SU\left( 2 \right) \times {T^2},
	\end{equation}
which has standard positive curvature.

	\end{itemize}

	\underline{\textbf{Case 2}: $V = 0$}
	
	Putting back $V = 0$ into  (\ref{linear-in-gamma}), one has
	\begin{equation}
		\left\{ \begin{array}{l}
		{g_{nk}}\left( {\nabla _m^A{t_I}^J} \right) - {g_{mk}}\left( {\nabla _n^A{t_I}^J} \right) = 0\\[0.5em]
		\left( {{\nabla _m}{F_{nk}}} \right) - \left( {{\nabla _n}{F_{mk}}} \right) = 0
		\end{array} \right..
	\end{equation}
These 2 condition implies covariant-constantness of $t_{IJ}$ and $F$:
	\begin{equation}
		\boxed{{\nabla _m^A{t_I}^J} = 0,\;{\nabla _k}{F_{mn}} = 0}\;.
	\end{equation}
In particular,
	\begin{equation}
		d * F = 0,\;\;dF = 0 \Leftrightarrow \Delta F = 0,
	\end{equation}
and 2nd/3rd Betti number is forced to be non-zero, if $F \ne 0$:
	\begin{equation}
		b^2 = b^3 \ge 1
	\end{equation}

	Compare the the terms {quadratic} in $\Gamma$, one obtains
	\begin{equation}
		\frac{1}{4}{R_{mnpq}}{\Gamma ^{pq}}{\delta _I}^J =  - 2{\left( {{t^2}} \right)_I}^J{\Gamma _{mn}} - \frac{1}{2}{F_{mp}}{F_{ns}}{\Gamma ^{ps}}{\delta _I}^J + \left[ {2{t_I}^J{F_{pm}}{\Gamma ^p}_n - \left( {m \leftrightarrow n} \right)} \right].
	\end{equation}

	The solutions are
	\begin{equation}
		{t_{IJ}} = 0\;\;\;{\rm{or }}\;\;\;F = 0.
	\end{equation}

	\underline{i) $t = 0$ while $F \ne 0, t=0$}:
	\begin{itemize}
		\item Riemann tensor
		\begin{equation}
			{R_{mnkl}} = {F_{ml}}{F_{nk}} - {F_{mk}}{F_{nl}}.
		\end{equation}
		
		Note that the expression satisfies interchange symmetry automatically, while the 1st Bianchi identity implies
		\begin{equation}
			{F_{m[l}}{F_{nk]}} - {F_{m[k}}{F_{nl]}} = 0 \Rightarrow F \wedge F = 0.
		\end{equation}
		
		\item Ricci tensor
		\begin{equation}
			Ric_{mn} = F_{mk} {F^k}_n.
		\end{equation}
.
		\item Scalar curvature
		\begin{equation}
			\mathcal{R} = - F_{mn} F^{mn},
		\end{equation}
which is {negative definite} if $F \ne 0$. Also note that $F$ is covariantly constant, hence $R_{mnnkl}$ is also {covariantly constant}. 
	
	Let's further constraint the form of curvature using the condition $F\wedge F = 0$. Noting that $F_{mn}$ is a $5\times 5$ antisymmetric matrix, we choose a coordinate where it takes block diagonal form:
	\begin{equation}
		F = {F_{12}}d{x^1} \wedge d{x^2} + {F_{34}}d{x^3} \wedge d{x^4}.
	\end{equation}
	
	Requiring that $F \wedge F = 0$ forces
	\begin{equation}
		{F_{12}}{F_{34}} = 0.
	\end{equation}
Assuming
	\begin{equation}
		F_{12} \ne 0,
	\end{equation}
with all other component zero, one arrives at a Riemann tensor with only one non-vanishing component:
	\begin{equation}
		{R_{1212}} =  - {\left( {{F_{12}}} \right)^2} < 0.
	\end{equation}
Combining with the fact that $F$ is parallel, this implies the 5-manifold $M$ should locally be product manifold
	\begin{equation}
		M = T^3 \times \mathbb{H}^2,
	\end{equation}
where $F = F_{12}dx^1\wedge dx^2$ serves as the volume form of $\mathbb{H}^2$.
	
	The metric of $M$ can be written as
	\begin{equation}
		d{s^2} = ds_{T^3}^2 + \frac{F_{12}}{y^2}\left( {d{x^2} + d{y^2}} \right).
	\end{equation}

	\end{itemize}

	\underline{ii) The case where $t\ne 0, F = 0$}
	\begin{itemize}
		\item Riemann tensor
		\begin{equation}
			{R_{mnkl}} = 2{\rm{tr}}\left( {{t^2}} \right)\left( {{g_{ml}}{g_{nk}} - {g_{mk}}{g_{nl}}} \right),
		\end{equation}
where interchange symmetry and first Bianchi identity are automatically satisfies.
		
		The second Bianchi identity forces ${\rm tr}(t^2)$ to be constant. The form of curvature implies that $M$ is a space of constant curvature, and therefore it must be locally isometric to $S^5$. This corresponds to the well-known fact that maximal number of solutions to the well-known Killing spinor equation can only be achieved on round $S^5$.
		
	\end{itemize}

	\section{Supersymmetric Theory for Vector Multiplet}\label{section-5}
	
	In section 4, we analyzed many properties of the proposed Killing spinor equation (\ref{Killing-equation} from supergravity, and discussed some necessary geometric conditions on the underlying manifold for solutions to exist.
	
	In this section, we propose a slightly generalized version of the supersymmetric theory for $\mathcal{N} = 1$ vector multiplet. It is not the most general one, as there are other known examples (constructed by dimensional reduction from $6d$, for instance) in recent literatures that does not completely fit in the following discussion. 
	
	Let us consider a simplified Killing spinor equation, where we set $F = 2V \equiv \mathcal{F}$ in (\ref{Killing-equation})
	\begin{equation}
		{D_m}{\xi _I} = {t_I}^J{\Gamma _m}{\xi _J} + \frac{1}{4}{\mathcal{F}^{pq}}{\Gamma _{mpq}}{\xi _I} + \frac{1}{2}{\mathcal{F}_{mn}}{\Gamma ^n}{\xi _I}.
		\label{SUSY-Killing}
	\end{equation}
$D_m$ contains Leve-civita connection, spin connection, gauge field $A_m$ from the vector multiplet and background $SU(2)$-gauge field ${A_I}^J$, depending on the objects it acts on. The change of notation to $\mathcal{F}_{mn}$ is to avoid confusion with the field strength of $\mathcal{N} = 1$ gauge field
	\begin{equation}
		{F_{mn}} \equiv {\nabla _m}{A_n} - {\nabla _n}{A_m} - i\left[ {{A_m},{A_n}} \right].
	\end{equation}

	We propose a supersymmetry transformation of $\mathcal{N} = 1$ vector multiplet with parameter $\xi$ being solution to the (\ref{SUSY-Killing}) is
	\begin{equation}
		\left\{ \begin{array}{l}
		{\delta _\xi	 }{A_m} = i{\epsilon ^{IJ}}\left( {{\xi _I}{\Gamma _m}{\lambda _J}} \right)\\[0.5em]
		{\delta _\xi }\phi  = i{\epsilon ^{IJ}}\left( {{\xi _I}{\lambda _J}} \right)\\[0.5em]
		\displaystyle{\delta _\xi }{\lambda _I} =  - \frac{1}{2}{F_{mn}}{\Gamma ^{mn}}{\xi _I} + \left( {{D_m}\phi } \right){\Gamma ^m}{\xi _I} + {\epsilon ^{JK}}{\xi _J}{D_{KI}} + 2\phi {t_I}^J{\xi _J} + \frac{1}{2}\phi {\mathcal{F} ^{pq}}{\Gamma _{pq}}{\xi _I}\\[0.5em]
		\displaystyle {\delta _\xi }{D_{IJ}} =  - i\left( {{\xi _I}{\Gamma ^m}{D_m}{\lambda _J}} \right) + \left[ {\phi ,\left( {{\xi _I}{\lambda _J}} \right)} \right] + i{t_I}^K\left( {{\xi _K}{\lambda _J}} \right) - \frac{i}{4}{\mathcal{F}^{pq}}\left( {{\xi _I}{\Gamma _{pq}}{\lambda _J}} \right) + \left( {I \leftrightarrow J} \right)
		\end{array} \right..
		\label{new_SUSY}
	\end{equation}
	
	Using previous results we obtain
	\begin{equation}
		d\kappa  = - 4\left( {t\Theta } \right) + 2s{V_V},\;\mathcal{F} =  - \frac{2}{s}\left( {t\Theta } \right) + \frac{2}{s}{\Omega ^ - } + {V_V},
	\end{equation}
with $V_V$ denoting the vertical part of field $V$.

	As discussed in an earlier remark, the above equation implies that $\kappa$ is a contact structure
	\begin{equation}
		\kappa  \wedge d\kappa  \wedge d\kappa  \ne 0.
	\end{equation}
Applying Weyl rescaling symmetry, one can eliminate $V_V$ and set $s = 1$. The Reeb vector field is then compatible with the contact structure $\kappa$:
	\begin{equation}
		\iota_R \kappa = 1, \iota_R d\kappa = 0.
	\end{equation}
Combining with the fact that $R$ is a Killing vector field, the structure $(\kappa, R, g)$ is actually a K-contact structure.
	
	For simplicity let us consider a special case where
	\begin{equation}
		\mathcal{F} = d\kappa,
	\end{equation}
namely $\Omega^- = 1/4 d\kappa$.

Then it is straight forward to prove that the following Lagrangian $S(\kappa, g)$ is invariant under (\ref{new_SUSY}):
	\begin{equation}
		\begin{array}{l}
		\displaystyle S = \int_M {\rm tr} \left [ {F \wedge *F - \kappa  \wedge F \wedge F - {d_A}\phi  \wedge *{d_A}\phi  - \frac{1}{2}{D_{IJ}}{D^{IJ}} + i{\lambda _I}{\slashed{D}_A}{\lambda ^I} - {\lambda _I}\left[ {\phi ,{\lambda ^I}} \right]} \right.  \\[0.8em]
		\displaystyle\;\;\;\;\;\;\; \left. - i{t^{IJ}}\left( {{\lambda _I}{\lambda _J}} \right) + 2\phi {t^{IJ}}{D_{IJ}} + \frac{i}{2}{\nabla_m \kappa_n}\left( {{\lambda _I}{\Gamma ^{mn}}{\lambda ^I}} \right) + 2 \phi F \wedge *d\kappa  + \frac{1}{4} \mathcal{R} \phi^2 \right]
		\end{array}.
		\label{new-Lag}
	\end{equation}
where $\mathcal{R}$ is the scalar curvature of the manifold.
	
	The detail proof will be presented in Appendix \ref{Appenidx-F}, but let us first make a few remarks here.
	
	As already mentioned, in the explicit form (\ref{new-Lag}) we took the choice to assume $\Omega^- = (1/4)d\kappa$, which is in fact a special case of a large family of supersymmetric theories in the following sense.
	
	Under supersymmetry (\ref{new_SUSY}) with $\xi$ satisfying (\ref{SUSY-Killing}) without imposing $\Omega^- = (1/4) d\kappa$, the Lagrangian without $\kappa\wedge F\wedge F$ has variation
	\begin{equation}
		\frac{i}{2}{{{\cal F}}_{mn}}{F_{pq}}\left( {{\xi _I}{\Gamma ^{mnpq}}{\lambda ^I}} \right)
	\end{equation}
	
	Such term can be identified in two ways. If we assume $\mathcal{F}$ is not only closed, but also exact
	\begin{equation}
		\mathcal{F} = d\mathcal{A} = \frac{1}{2}d\kappa  + 2{\Omega ^ - }
	\end{equation}
for some 1-form $\mathcal{A}$, then the term can be identified as variation of
	\begin{equation}
		\mathcal{A}  \wedge F \wedge F
	\end{equation}
In such case, the theory is specified by $\kappa$ and $\mathcal{A}$.
	
	However, if we do not assume anything of $\mathcal{F}$, then the term can also be identified as variation of
	\begin{equation}
		{{\cal F}} \wedge \left( {A \wedge dA + \frac{2}{3}A \wedge A \wedge A} \right)
	\end{equation}
In such case, the theory is specified by nowhere-vanishing 1-form $\kappa$ and a closed anti-self-dual 2-form $\Omega^-$, although the gauge invariance is not nicely manifested.
	
	Following an analysis similar to that in \cite{Hosomichi:2012fk}, one can add to the Lagrangian (\ref{new-Lag}) a $\delta$-exact term $\delta_\xi V$ with
	\begin{equation}
		V = {\rm{tr}}\left( {{{\left( {{\delta _\xi }\lambda } \right)}^\dag }\lambda } \right).
	\end{equation}
Then the localization locus is
	\begin{equation}
		F_H^ -  = \phi d\kappa ,\;\;{\iota _R}F = 0,\;\;{d_A}\phi  = 0,\;\;{D_{IJ}} + 2\phi {t_{IJ}} = 0
	\end{equation}
For general $\Omega^-$, the first equation will take a more general form
	\begin{equation}
		F_H^ -  = \phi d\kappa  + \phi \Omega _H^ - .
	\end{equation}
This localization locus is a generalization of the contact instanton in \cite{Kallen:2012fk}.

It would be interesting to perform a complete localization for the theory (\ref{new-Lag}) with the above localization locus, which we leave for future study.

\section{Discussion}

	So far we have obtained many constraints on geometry of $M$ imposed by the existence of supercharges. For 1 pair of supercharges, generically $M$ must be almost contact manifold, and using the compatible connection, the Killing spinor equation can be simplified to a compact form. We also discussed a few interesting cases related to product manifold.d, which leads to special foliation and reduction to known 4-dimensional Killing spinor equations. The presence of 2 pairs supercharges with 2 additional assumptions restricts the isometry algebra of $M$, forcing $M$ to be $S^3$ or $T^3$-fibration over Riemann surfaces. The presence of 4 pairs of supercharges allows for only 3 major possibilities, where the corresponding topologies and geometries are basically fixed.
	
	There are several problems that are interesting to explore further.
	
	(1) We obtained necessary conditions for supercharges to exist, but not sufficient conditions. In 3 dimension\cite{Closset:2012zr}, the general solution to Killing spinor equation is obtained from the special coordinate, which requires some integrability of the almost contact structure. However, we do not have such integrability for the almost contact structure we defined, partly because the definition involves auxiliary field $t_{IJ}$, and the differential property of $t_{IJ}$ is not known at priori. Moreover, in the extreme case where $t_{IJ} = 0$, it is not obvious that $M$ is still a almost contact manifold. Perhaps it is possible to define almost contact structure of $M$ without referring to $t_{IJ}$. 
	
	(2) We partially solved the auxiliary fields, but not all: gauge field $A$ and $t_{IJ}$ are entangled together. If $t_{IJ}$ and $A$ could be solved in terms of pure bilinears separately, the first problem above can also be solved. 
	
	(3) In the discussions, we made a few assumptions and simplifications. For examples, we did not study all possible bilinears formed by all solutions, but focused on those formed by the representatives from each pair. One should be able to obtain more information of $M$ by taking into account all of them. Also, to simplify computation we assumed $\Omega^- = 0$ in some discussions. It is straight-forward and interesting to reinstate general $\Omega^-$, and understand its role in the almost contact metric structure.
	
	(4) We start from Zucker's off-shell supergravity\cite{Zucker:2000uq}. However, it is not coupled to matter fields, and hence one would not automatically obtain any supersymmetric theory for matter multiplets. Our analysis, in this sense, is far from enough to obtain a complete picture. A next step one could try is to start from known 5-dimensional off-shell supergravity coupled with matter and take the rigid limit. For instance, one can start with $\mathcal{N} = 1$ supergravity coupled to Yang-Mills matters in \cite{Kugo:fk}, which was considered in \cite{Hosomichi:2012fk}. After turning on auxiliary fields $t_{IJ}$ and $V_{mn}$, the Killing spinor equation involved is then
	\begin{equation}
		{\nabla _m}{\xi _I} = {t_I}^J{\Gamma _m}{\xi _J} + \frac{1}{2}{V_{mpq}}{\Gamma ^{pq}}{\xi _I},
	\end{equation}
which is a special case of our more general equation.

\section{Examples}
	
	In this section, we present simple explicit examples that illustrate some of the discussion before, by solving Killing spinor equations on selected manifolds and determining the auxiliary fields. 
	
\subsection{$M = S^1 \times S^4$}

	In earlier discussion, we discussed the possibility of having $M = S^1 \times N$ with $N$ a 4d Quaternion-K\"ahler manifold. In this section, we consider the case where $N = S^4$.
	
	Denote the coordinate along $S^1$ to be $\theta$, $x^\mu$ are stereo-projection coordinates on $S^4$. The metric of $S^1 \times S^4$ is simply
	\begin{equation}
		d{s^2} = d{\theta ^2} + \frac{{{\delta _{\mu \nu }}d{x^\mu }d{x^\nu }}}{{{{\left( {1 + {r^2}} \right)}^2}}}
	\end{equation}
with function ${r^2} = \sum\limits_{\mu  = 1}^4 {{{\left( {{x^\mu }} \right)}^2}} $
	
	As discussed before, we partially fix the auxiliary fields
	\begin{equation}
		F = 0,\;V = t\Theta 
	\end{equation}
However, non-zero $t\Theta$ will generate globally defined almost complex structure on $S^4$, which we already know  does not exist, hence we can set $t = 0$ and $V = 0$. The only auxiliary fields allowed is thus $SU(2)$ gauge field $A$.

	The Killing spinor equation (\ref{Rewritten-Killing}) now reads
	\begin{equation}
		\left\{ \begin{array}{l}
		{\partial _\theta }{\xi _I} = {({{\hat A}_\theta })_I}^J{\xi _J}\\[0.5em]
		{\nabla _\mu }{\xi _I} = {({{\hat A}_\mu })_I}^J{\xi _J}
		\end{array} \right.
	\end{equation}
	
	The gauge field $A_\mu$ is determined by the a choice of Quaternion-K\"ahler structure on $S^4$. Denoting
	\begin{equation}
		{z_1} \equiv {x^1} + i{x^2},\;{z_2} \equiv {x^3} + i{x^4}
	\end{equation}
one can define locally 3 almost complex structures as the basis,
	\begin{equation}
		\left\{ \begin{array}{l}
		\displaystyle {J^1} = \left( {\frac{\partial }{{\partial \overline {{z_1}} }} \otimes d{z_2} - \frac{\partial }{{\partial \overline {{z_2}} }} \otimes d{z_1}} \right) + h.c.\\[0.5em]
		\displaystyle{J^2} = \frac{1}{i}\left( {\frac{\partial }{{\partial \overline {{z_1}} }} \otimes d{z_2} - \frac{\partial }{{\partial \overline {{z_2}} }} \otimes d{z_1}} \right) + h.c.\\[0.5em]
		\displaystyle{J^3} = i\frac{\partial }{{\partial {z_i}}} \otimes d{z_i} - i\frac{\partial }{{\partial \overline {{z_i}} }} \otimes d\overline {{z_i}} 
		\end{array} \right.
	\end{equation}
and determine the gauge field using (\ref{gauge-field}).
	
	We choose the Gamma matrices to be
	\begin{equation}
		{\Gamma ^i} = {\sigma ^i} \otimes {\sigma ^2},\;{\Gamma ^4} = I \otimes {\sigma ^1},\;{\Gamma ^5} = I \otimes {\sigma _3},\;C = {\Gamma ^{13}}
	\end{equation}
and the obvious vielbein
	\begin{equation}
		{e^5} = - d\theta ,\;{e^a} = \frac{1}{{1 + {r^2}}}\delta _\mu ^ad{x^\mu }
	\end{equation}
solution is given as
	\begin{equation}
		{\xi _1} = {e^{i\int {{A_\theta }d\theta } }}{\chi _ + } \otimes {\chi _ - },\;{\xi _2} = - {e^{ - i\int {{A_\theta }d\theta } }}{\chi _ - } \otimes {\chi _ - }
	\end{equation}

\subsection{$M = S^2 \times S^3$}

	Consider $S^3$ as a $U(1)$ bundle over $S^2$. Let $S^3$ be embedded into $\mathbb{C}^2$,
	\begin{equation}
		{S^3} = \left\{ {{{\left| {{z_1}} \right|}^2} + {{\left| {{z_2}} \right|}^2} = 1|\left( {{z_i}} \right) \in {\mathbb{C}^2}} \right\}
	\end{equation}
	
	Similarly define
	\begin{equation}
		{z_2} = \rho {e^{i\theta }},\;z \equiv \frac{{{z_1}}}{{{z_2}}} \Rightarrow {\left. {{\rho ^2}} \right|_{{S^3}}} = \frac{1}{{1 + {{\left| z \right|}^2}}}, \;{z_1} = z{z_2} = \rho {e^{i\theta }}z
	\end{equation}
and hence the induced round metric on $S^3$ can be written as
	\begin{equation}
		\begin{array}{l}
		\displaystyle d{s^2} = d{z_1}d\overline {{z_1}}  + d{z_2}d\overline {{z_2}}  = {\left[ {d\theta  + i\frac{{zd\bar z - \bar zdz}}{{2\left( {1 + {{\left| z \right|}^2}} \right)}}} \right]^2} + \frac{{dzd\bar z}}{{{{\left( {1 + {{\left| z \right|}^2}} \right)}^2}}}\\[0.5em]
		\displaystyle\;\;\;\; = {\left( {d\theta  + a} \right)^2} + g^1
		\end{array}
	\end{equation}
where
	\begin{equation}
		g^1 = \frac{{dzd\bar z}}{{{{\left( {1 + {{\left| z \right|}^2}} \right)}^2}}}	
	\end{equation}
is the metric on $\mathbb{C}P^1 = S^2$ with radius $1/2$. In coordinate,
	\begin{equation}
		{g^1_{z\bar z}}= {g^1_{\bar zz}} = \frac{1}{{2{{\left( {1 + {{\left| z \right|}^2}} \right)}^2}}}  = \frac{1}{2} \partial_z \partial_{\bar z} \ln (1 + \left | z \right |^2) \equiv \partial_z \partial_{\bar z} K
	\end{equation}
and
	\begin{equation}
		.	
	\end{equation}

	The vector field $R \equiv \partial_\theta$ is a Killing vector field, and its dual is $\kappa = d\theta +A$, such that ${\iota _R}\kappa  = 1$.
	
	Define the frame on $S^3$ to be
	\begin{equation}
		e^3 \equiv {e^\theta } = \kappa ,\;{e^1} = \frac{{{\mathop{\rm Re}\nolimits} dz}}{{1 + {{\left| z \right|}^2}}},\;\;{e^2} = \frac{{{\mathop{\rm Im}\nolimits} dz}}{{1 + {{\left| z \right|}^2}}}
	\end{equation}
s.t.
	\begin{equation}
		g = {e^\theta }{e^\theta } + {e^1}{e^1} + {e^2}{e^2}
	\end{equation}
then it is obvious that
	\begin{equation}
		{\omega _{\theta ab}} = 0,  a,b = \theta, 1,2
	\end{equation}
from the fact 
	\begin{equation}
		de^\theta \sim \frac{{idz \wedge d\bar z}}{{{{\left( {1 + {{\left| z \right|}^2}} \right)}^2}}}
	\end{equation}
	
	The base manifold $S^2 \times S^2$ is complex, with natural complex structure and K\"ahler form. Setting the radius of the stand-alone $S^2$ to be $l$, with local complex coordinate $w$, the metric of $S^3 \times S^2$ reads
	\begin{equation}
		g = {\left( {d\theta  + a} \right)^2} + \frac{{dzd\bar z}}{{{{\left( {1 + {{\left| z \right|}^2}} \right)}^2}}} + \frac{{4{l^2}dwd\bar w}}{{{{\left( {1 + {{\left| w \right|}^2}} \right)}^2}}}
	\end{equation}
with K\"ahler form on base manifold
	\begin{equation}
		\omega  = \frac{{i dz \wedge d\bar z}}{{{{ 2 \left( {1 + {{\left| z \right|}^2}} \right)}^2}}} + \frac{{i4{l^2}dw \wedge d\bar w}}{2{{{\left( {1 + {{\left| w \right|}^2}} \right)}^2}}}
	\end{equation}
or in components
	\begin{equation}
		{\omega _{z\bar z}} =  - {\omega _{\bar zz}} = \frac{i}{2 \left ( 1+ \left| z \right|^2 \right)^2} ,\; {\omega _{w\bar w}} =  - {\omega _{\bar ww}} = i{g_{w\bar w}}  = \frac{il}{2 \left( 1+ \left | z \right )^2 \right )^2}
	\end{equation}
	
	The 2 complex structures on both $\mathbb{C}P^1$ can form linear combination
	\begin{equation}
		\varphi_\pm  \equiv {J_1} \pm {J_2}
	\end{equation}
which satisfies
	\begin{equation}
		{\varphi_\pm ^2} =  - 1 + R \otimes \kappa 
	\end{equation}
	
	Let us now construct the auxiliary fields. We choose $t_{IJ}$ such that $\displaystyle {\rm{tr}}\left( {{t^2}} \right) =  - \frac{1}{2}$, and therefore
	\begin{equation}
		4{\left( {t\Theta } \right)^2}\sim - 1 + ...
	\end{equation}
We identify a combination of the 2 complex structures on 2 $\mathbb{C}P^1$ as $t\Theta$. Recall that $t \Theta$ also satisfies ${\iota _R}*\left( {t\Theta } \right) =  - \left( {t\Theta } \right)$, hence we identify
	\begin{equation}
		{\varphi _ - }\sim2\left( {t\Theta } \right)
	\end{equation}
or a 2-form equation
	\begin{equation}
		2\left( {t\Theta } \right) = \frac{{idz \wedge d\bar z}}{{2{{\left( {1 + {{\left| z \right|}^2}} \right)}^2}}} - \frac{{i4{l^2}dw \wedge d\bar w}}{{2{{\left( {1 + {{\left| w \right|}^2}} \right)}^2}}}
	\end{equation}
	
	Then we obtain $F$ and $V$:
	\begin{equation}
		F = \frac{1}{2}d\kappa  = \frac{{idz \wedge d\bar z}}{{2{{\left( {1 + {{\left| z \right|}^2}} \right)}^2}}}
	\end{equation}
and
	\begin{equation}
		V  = t\Theta  = \frac{{idz \wedge d\bar z}}{{4{{\left( {1 + {{\left| z \right|}^2}} \right)}^2}}} - \frac{{i{l^2}dw \wedge d\bar w}}{{{{\left( {1 + {{\left| w \right|}^2}} \right)}^2}}}
	\end{equation}
	
	With these auxiliary fields, one can solve the Killing spinor equation
	\begin{equation}
		{{\hat \nabla }_m}{{\hat \xi }_I} = {{( {{{\hat A}_m}\;} )_{}}_I}^J{{\hat \xi }_J}
	\end{equation}
Denote $\alpha = w, \bar w$, and $\mu,\nu = z, \bar z$, we have
	\begin{equation}
		\left\{ {\begin{array}{*{20}{l}}
		{{\nabla _\alpha }{\xi _I} = {{\left( {{A_\alpha }} \right)}_I}^J{\xi _J}}\\[0.5em]
		\displaystyle {{\nabla _\mu }{\xi _I} - \frac{1}{2}\left( {{\nabla _\mu }{R_\nu }} \right){\Gamma ^\nu }		{\xi _I} = {{( {{\hat A_\mu }} )}_I}^J{\xi _J} }\\[0.5em]
		{{\nabla _\theta }{\xi _I} = {{( {{\hat A_\theta }} )}_I}^J{\xi _J} }
		\end{array}} \right.
	\end{equation}
where
	\begin{equation}
		{R_\theta} = 1,\; {R_z} = \frac{1}{2}\frac{{ - i\bar z}}{{1 + {{\left| z \right|}^2}}} = -i\partial_z K,\;{R_{\bar z}} = \frac{1}{2}\frac{{iz}}{{1 + {{\left| z \right|}^2}}}= i\bar \partial _{\bar z }K
	\end{equation}
and we used
	\begin{equation}
		{\nabla _\mu }{R_\theta } = {\nabla _\theta }{R_\theta } = 0
	\end{equation}
	
	Choosing gauge field to be $(A_m)_I^J = (A_m)(\sigma_3)_I^J$, 
	\begin{equation}
		i{A_z} = \frac{{\bar z}}{{4\left( {1 + {{\left| z \right|}^2}} \right)}},\;\;i{A_{\bar z}} =  - \frac{z}{{4\left( {1 + {{\left| z \right|}^2}} \right)}}, A_\theta = - \frac{1}{4}
	\end{equation}
and representation of Gamma matrices
	\begin{equation}
		{\Gamma _{w,\bar w}}\sim{\sigma _{1,2}} \otimes 1,\;{\Gamma _{z,\bar z,\theta }}\sim{\sigma _3} \otimes {\sigma _{1,2,3}}
	\end{equation}
one obtains the chiral solution ($\xi_2$ is obtained from symplectic Majorana condition)
	\begin{equation}
		\displaystyle \xi_1 = {e^{ - \frac{i}{4}\theta }}{\chi _ + } \otimes {\chi _ + }
	\end{equation}
	
	The calculation can be easily generalized to $M = S^3 \times \Sigma$ for Riemann surface $\Sigma$.

\section*{Acknowledgments}

	The author of the paper would like to thank his advisor Martin Ro\v{c}{e}k for guidance and inspiring discussions. I also thank Maxim Zabzine and Guido Festuccia for carefully reading the draft and pointing out several points that were not stated clearly, as well as many other suggestions. The author also had helpful discussions with fellow students Jun Nian and Xinyu Zhang. The author would like thank NSF grant no. PHY-1316617 for partial support.

\begin{appendices}

\section{Gamma matrices and Fierz identities}\label{Appendix-A}
	We denote the 5d Gamma matrices as $\Gamma^m$ with defining anti-commutation relation
	\begin{equation}
		\left\{ {{\Gamma ^m},{\Gamma ^n}} \right\} = 2{g^{mn}}
	\end{equation}
We require them to be {Hermitian}
	\begin{equation}
		{\left( {{\Gamma ^m}} \right)^\dag } = {\Gamma ^m}
	\end{equation}
	
	Also we have {charge conjugation matrix} $C=C_{+}$
	\begin{equation}
		C{\Gamma ^m}{C^{ - 1}} = {\left( {{\Gamma ^m}} \right)^T} = \overline {{\Gamma ^m}} 
	\end{equation}
	
	These matrices have the following symmetry properties:
	\begin{equation}
		{C_{\alpha \beta }} =  - {C_{\beta \alpha }},\;\;{\left( {C{\Gamma _m}} \right)_{\alpha \beta }} =  - {\left( {C{\Gamma _m}} \right)_{\beta \alpha }}
	\end{equation}
	\begin{equation}
		{\left( {C{\Gamma _{mn}}} \right)_{\alpha \beta }} = {\left( {C{\Gamma _{mn}}} \right)_{\beta \alpha }},\;\;{\left( {C{\Gamma _{lmn}}} \right)_{\alpha \beta }} = {\left( {C{\Gamma _{lmn}}} \right)_{\beta \alpha }}
	\end{equation}
	and complex conjugation properties
	\begin{equation}
		\sum\limits_\beta  {\overline {{C_{\alpha \beta }}} {C_{\beta \gamma }}}  = - {\delta ^\alpha }_\gamma ,\;\overline {{{\left( {{\Gamma ^m}} \right)}^\alpha }_\beta }  = {\left( {{\Gamma ^m}} \right)^\beta }_\alpha 
	\end{equation}
	
	The symmetry properties of $C \Gamma$ results in symmetry properties of bilinears of spinors:
	\begin{equation}
		\begin{array}{l}
		\left( {{\xi _1}{\xi _2}} \right) =  - \left( {{\xi _2}{\xi _1}} \right),\;\;\left( {{\xi _1}{\Gamma _m}{\xi _2}} \right) =  - \left( {{\xi _2}{\Gamma _m}{\xi _1}} \right)\\[0.5em]
		\left( {{\xi _1}{\Gamma _{mn}}{\xi _2}} \right) = \left( {{\xi _2}{\Gamma _{mn}}{\xi _1}} \right),\;\;\left( {{\xi _1}{\Gamma _{lmn}}{\xi _2}} \right) = \left( {{\xi _2}{\Gamma _{lmn}}{\xi _1}} \right)
		\end{array}	
	\end{equation}

	In $5d$, we have 
	\begin{equation}
		{\Gamma ^1}...{\Gamma ^5} = 1  \Leftrightarrow {\Gamma ^{abcde}} = {\epsilon ^{abcde}}
	\end{equation}
following from the fact that $\left[ {{\Gamma ^1}...{\Gamma ^5},{\Gamma ^a}} \right] = 0$ and Schur lemma.

	This fact has the following duality consequence:
	
	\textbf{Proposition}
	\begin{equation}
		{\Gamma ^{abcd}} = {\epsilon ^{abcde}}{\Gamma ^e} \Leftrightarrow \frac{1}{{4!}}{\epsilon ^{abcde}}{\Gamma _{abcd}} = {\Gamma ^e}
	\end{equation}
	and
	\begin{equation}
		\frac{1}{{3!}}{\epsilon ^{abcde}}{\Gamma ^{abc}} =  - {\Gamma ^{de}} \Leftrightarrow \frac{1}{{2!}}{\epsilon ^{abcde}}{\Gamma ^{ab}} =  - {\Gamma ^{cde}}
	\end{equation}
	
	The Hodge star operator associated with metric $g_{mn}$ is defined as
	\begin{equation}
		*d{x^{{i_1}}} \wedge ... \wedge d{x^{{i_p}}} = \frac{{\sqrt g }}{{\left( {n - p} \right)!}}{\epsilon ^{{i_1}...{i_1}}}_{{j_1}...{j_{n - p}}}d{x^{{j_1}}} \wedge ... \wedge d{x^{{j_{n - p}}}}
		\label{Hodge_1}
	\end{equation}
	or equivalently for $\displaystyle {\omega _{\left( p \right)}} \equiv \frac{1}{{p!}}{\omega _{{i_1}...{i_p}}}d{x^{{i_1}}} \wedge ... \wedge d{x^{{i_p}}}$
	\begin{equation}
		{\left( {*\omega } \right)_{{j_1}...{j_{n - p}}}} = \frac{\sqrt{g}}{{p!}}{\epsilon ^{{i_1}...{i_p}}}_{{j_1}...{j_{n - p}}}{\omega _{{i_1}...{i_p}}}
		\label{Hodge_2}
	\end{equation}
	\begin{equation}
		*\omega  = \frac{{\sqrt g }}{{p!\left( {n - p} \right)!}}{\epsilon ^{{i_1}...{i_p}}}_{{j_1}...{j_{n - p}}}{\omega _{{i_1}...{i_p}}}d{x^{{j_1}}} \wedge ... \wedge d{x^{{j_{n - p}}}}
	\end{equation}
	
	Let us define $p$-forms $\Theta_{(p)}$ as
	\begin{equation}
		\Theta _{\left( p \right)}^{IJ} = \frac{1}{{p!}}\left( {{\xi ^I}{\Gamma _{{i_1}...{i_p}}}{\xi ^J}} \right)d{x^{{i_1}}} \wedge ... \wedge d{x^{{i_p}}}
	\end{equation}
They satisfy
	\begin{equation}
		*\Theta _{\left( 2 \right)}^{IJ} =  - \Theta _{\left( 3 \right)}^{IJ}
	\end{equation}
	\begin{equation}
		*\Theta _{\left( 1 \right)}^{IJ} = \Theta _{\left( 4 \right)}^{IJ}
	\end{equation}
In components, they are
	\begin{equation}
		\left( {{\xi ^I}{\Gamma _{lmn}}{\xi ^J}} \right) = - {\left( {*{\Theta^{IJ} _{\left( 2 \right)}}} \right)_{lmn}},\;\;\left( {{\xi ^I}{\Gamma _{mnpq}}{\xi ^J}} \right) = *{\left( {\Theta _{\left( 1 \right)}^{IJ}} \right)_{mnpq}} = \frac{1}{2}\epsilon^{IJ}{\left( {*\kappa } \right)_{mnpq}}
	\end{equation}

	\vspace{20pt}

	For any 2 spinors $\xi_1$ and $\xi_2$, we define their inner product as a complex number:
	\begin{equation}
		\left( {{\xi _1}{\xi _2}} \right) \equiv \xi _1^\alpha {C_{\alpha \beta }}\xi _2^\beta 
	\end{equation}
	with symmetry properties
	\begin{equation}
		\left( {{\xi _1}{\xi _2}} \right) = - \left( {{\xi _2}{\xi _1}} \right),\;\;\left( {{\xi _1}{\Gamma _m}{\xi _2}} \right) = - \left( {{\xi _2}{\Gamma _m}{\xi _1}} \right)
	\end{equation}
	\begin{equation}
		\left( {{\xi _1}{\Gamma _{mn}}{\xi _2}} \right) =   \left( {{\xi _2}{\Gamma _{mn}}{\xi _1}} \right),\;\;\left( {{\xi _1}{\Gamma _{lmn}}{\xi _2}} \right) =    \left( {{\xi _2}{\Gamma _{lmn}}{\xi _1}} \right)
	\end{equation}
	and
	\begin{equation}
		\left( {\left( {{\Gamma _m}{\xi _1}} \right){\xi _2}} \right) = \left( {{\xi _1}{\Gamma _m}{\xi _2}} \right)
	\end{equation}
	
	{Fierz identity} \cite{Hosomichi:2012fk}: for any 3 Grassmann even spinors $\left( {{\xi _1},{\xi _2},{\xi _3}} \right)$, one has Fierz identity
	\begin{equation}
		{\xi _1}\left( {{\xi _2}{\xi _3}} \right) =  \frac{1}{4}{\xi _3}\left( {{\xi _2}{\xi _1}} \right) + \frac{1}{4}{\Gamma ^m}{\xi _3}\left( {{\xi _2}{\Gamma _m}{\xi _1}} \right) - \frac{1}{8}{\Gamma ^{mn}}{\xi _3}\left( {{\xi _2}{\Gamma _{mn}}{\xi _1}} \right)
		\label{Fierz_1}
	\end{equation}
It immediately follows from the above Fierz identity that
	\begin{equation}
		{\Gamma ^m}{\xi _1}\left( {{\xi _2}{\Gamma _m}{\xi _3}} \right) + {\xi _1}\left( {{\xi _2}{\xi _3}} \right) = 2{\xi _3}\left( {{\xi _2}{\xi _1}} \right) - 2{\xi _2}\left( {{\xi _3}{\xi _1}} \right)
		\label{BosFierz_2}
	\end{equation}
	\begin{equation}
		{\xi _1}\left( {{\xi _2}{\xi _3}} \right) + {\xi _2}\left( {{\xi _1}{\xi _3}} \right) =  - \frac{1}{4}{\Gamma ^{pq}}{\xi _3}\left( {{\xi _2}{\Gamma _{pq}}{\xi _1}} \right)
		\label{BosFierz_3}
	\end{equation}

\section{$SU(2)$ indices and some notations}\label{Appendix-B}

	In the main text we frequently have to deal with the $SU(2)$ indices. 
	
	The $SU(2)$-invariant tensor $\epsilon$ is defined as ${\epsilon ^{12}} = {\epsilon _{21}} = 1$, with contraction
	\begin{equation}
		{\epsilon ^I}_K = {\epsilon ^{IJ}}{\epsilon _{JK}} =  - {\epsilon ^{IJ}}{\epsilon _{KJ}} =  - {\epsilon _K}^I = {\delta ^I}_K \Rightarrow {\epsilon ^{IJ}}{\epsilon _{IJ}} =  - 2	
	\end{equation}
	and raising/lowering rules
	\begin{equation}
		{X^I} = {\epsilon ^{IJ}}{X_J} \Leftrightarrow {X_I} = {\epsilon _{IJ}}{X^J}
	\end{equation}

	With this "metric", we define for any 2 triplets of functions $X^{IJ}$ and $Y^{IJ}$ a product in a natural way:
	\begin{equation}
		{\left( {XY} \right)^{IJ}} \equiv {\epsilon _{LK}}{X^{IK}}{Y^{LJ}} = {X^I}_K{Y^{KJ}}
		\label{triplet-product}
	\end{equation}
	
	Note that this product has the following symmetry:
	\begin{equation}
		{\left( {XY} \right)^{IJ}} =  - {\left( {YX} \right)^{JI}}
		\label{triplet-product-symm}
	\end{equation}
	and in particular
	\begin{equation}
		{\left( {{X^2}} \right)^{IJ}} =  - {\left( {{X^2}} \right)^{JI}} = \frac{1}{2}{\rm tr}(X^2) \epsilon^{IJ}
		\label{triplet-product-symm-2}
	\end{equation}
	where we define the trace for triplet products:
	\begin{equation}
		{\rm{tr}}\left( {XY} \right)  \equiv {X_I}^J {Y_J}^I = - X_{IJ} Y^{IJ}
	\end{equation}
with cyclic symmetry
	\begin{equation}
		{\rm{tr}}\left( {XY} \right) = {\rm{tr}}\left( {YX} \right)
		\label{cyclic}
	\end{equation}
	
	As an example, when ${X_I}^J = \displaystyle \frac{i}{2}{\left( {{\sigma _3}} \right)_I}^J$
	\begin{equation}
		{\rm tr}{X^2} =  - \frac{1}{2}
	\end{equation}
Note that if non-zero quantity $X_{IJ}$ satisfies reality condition
	\begin{equation}
		\overline {{X_{IJ}}}  = {\epsilon ^{II'}}{\epsilon ^{JJ'}}{X_{I'J'}}
	\end{equation}
then the trace is negative definite
	\begin{equation}
		{\rm{tr}}\left( {{X^2}} \right) < 0
	\end{equation}
	
	For objects of direct sum of triplet and singlet,
	\begin{equation}
		{X_{IJ}} \equiv {\hat X_{IJ}} - \frac{1}{2}{\epsilon _{IJ}}X,\;{X^{IJ}} \equiv {{\hat X}^{IJ}} + \frac{1}{2}{\epsilon ^{IJ}}X
	\end{equation}
with
	\begin{equation}
		X = {\epsilon ^{IJ}}{X_{IJ}} =  - {\epsilon _{IJ}}{X^{IJ}}
	\end{equation}

\section{Differential Geometry}\label{Appendix-C}

	In the main text, we denote Levi-civita connection on $M$ as $\nabla$:
	\begin{equation}
		\nabla g = 0
	\end{equation}
with connection coefficients
	\begin{equation}
		{\Gamma ^k}_{mn} = \frac{1}{2}{g^{kl}}\left( {\frac{{\partial {g_{ml}}}}{{\partial {x^n}}} + \frac{{\partial {g_{nl}}}}{{\partial {x^m}}} - \frac{{\partial {g_{mn}}}}{{\partial {x^l}}}} \right)
	\end{equation}
and curvature tensor
	\begin{equation}
		\left[ {{\nabla _m},{\nabla _n}} \right]{X^k} = {R_{mnl}}^k{X^l}
	\end{equation}
	
	Ricci tensor is defined as
	\begin{equation}
		Ri{c_{mn}} = {R_{mkn}}^k
	\end{equation}
	
and covariant derivative on spinor is 
	\begin{equation}
		{\nabla _m}\psi  = {\partial _m}\psi  - \frac{1}{4}{\omega _{mab}}{\Gamma ^{ab}}\psi 
	\end{equation}
where the spin connection is defined as
	\begin{equation}
		{\omega _{ma}}{^b} \equiv {e^b}\left( {{\nabla _m}{e_a}} \right) = {e^b}_n{\nabla _m}{e_a}^n = {e^b}_n{\partial _m}e_a^n + {\Gamma ^b}_{ma}
	\end{equation}

	Lie derivative for (1,1), (0,2) tensor are defined as
	\begin{equation}
		{\mathcal{L}_X}{T^m}_n = {X^k}{\nabla _k}{T^m}_n - \left( {{\nabla _k}{X^m}} \right){T^k}_n + \left( {{\nabla _n}{X^k}} \right){T^m}_k
	\end{equation}
	\begin{equation}
		{\mathcal{L}_X}{T_{mn}} = {X^k}{\nabla _k}{T}_{mn} + \left( {{\nabla _m}{X^k}} \right){T_{kn}} + \left( {{\nabla _n}{X^k}} \right){T_{mk}}
	\end{equation}
with the obvious relation
	\begin{equation}
		{\mathcal{L}_X}{T_{mn}} = \left( {{\nabla _m}{X_l} + {\nabla _l}{X_m}} \right){T^l}_n + {g_{ml}}{\mathcal{L}_X}{T^l}_n
	\end{equation}

\section{Contact,  Almost Contact Structure and Compatible Connection}\label{Appendix-D}

	In this appendix,we will introduce necessary background on contact geometry. It is the odd dimensional analog of symplectic geometry in even dimension. Compared to its even dimensional sister, contact geometry is much less studied. However, there are interesting developments in the past few years, on the existence and classification of contact structures ,as well as Sasaki-Einstein structures.
	
	\vspace{20pt}
	
	The Euler number of any odd dimensional manifold $M^{2n+1}$ is zero, therefore one can have nowhere vanishing vector fields. Contact geometry and almost contact geometry studies the behavior of these vector fields, or their corresponding hyperplane fields.
	
	Suppose one has a nowhere vanishing 1-form $\kappa$ on $d= 2n+1$-dimensional $M$. $\kappa$ singles out a rank-$2n$ vector bundle $TM_H$ as a sub-bundle of $TM$, such that at $p \in M$, $T_pM_H = {\rm ker}\kappa_p$. The sub-bundle $TM_H$ is usually called oriented hyperplane field. As $\kappa$ is nowhere-vanishing, the quotient line bundle $TM_V \equiv TM/ TM_H$ is trivial. Let us call $TM_H$ horizontal vector bundle, and $TM_V$ as vertical bundle. Note that specifying a oriented hyperplane field is equivalent to specifying $\kappa$ up to any nowhere vanishing real function factor.
	
	Recall that $TM_H$ is integrable if and only if
	\begin{equation}
		d\kappa \wedge \kappa = 0 \Leftrightarrow d\kappa = \kappa \wedge (...)
	\end{equation}
and $M$ is locally foliated by the integral manifold.

	Contact structure, however, sits in the opposite extreme. It is completely non-integrable in the sense that the hyperplane fields cannot be smoothly patched together to be submanifolds. $\kappa$ satisfies the non-degenerate condition
	\begin{equation}
		\kappa  \wedge {\left( {d\kappa } \right)^n} \ne 0
	\end{equation}
which remains true when $\kappa$ is rescaled by nowhere-vanishing function. This condition implies $d\kappa$ is of maximal rank $2n$, but $\kappa$ and $d\kappa$ do not have common zero eigenvector. Therefore, at $\forall p \in M$ there exist a line in $T_p M$ such that on the line $d\kappa = 0$ but not $\kappa$. Then one can choose a vector along this line at each point such that the resulting vector field $R$ satisfies
	\begin{equation}
		\kappa \left( R \right) = 1,\;{\iota _R}d\kappa  = 0
	\end{equation}
Such vector field $R$ is called Reeb vector field.

	In low dimensions, contact structure is ubiquitous. In 3 dimension, every orientable manifold admits a contact structure, thanks to Thurston's geometrization. In 5 dimension, contact structure exists on manifolds with vanishing third integral Stiefel-Whitney class\cite{Hamilton:2010aa}. However, it is not clear if similar holds true in higher dimension. We will comment on this after we discuss the almost contact structure.

	For any contact structure, one can associate a metric $g$ such that
	\begin{equation}
		g\left( {R, \cdot } \right) = \kappa \left(  \cdot  \right)
	\end{equation}
and consequently $\kappa$ and $R$ have unit norm. Actually, there are infinitely many such associate metrics compatible with the contact structure. In this case, $(\kappa, R, g)$ is called contact metric structure.

	Another kind of similar structure exists on contact manifolds is called almost contact structure.  It is defined by a triplet $(R^m, \kappa_m, {\varphi_m}^n)$ satisfying
	\begin{equation}
		\left\{ \begin{array}{l}
		{R^m}{\kappa _m} = 1\\[0.5em]
		{\varphi _m}^n{R^m} = 0\\[0.5em]
		{\varphi _m}^n{\kappa _n} = 0\\[0.5em]
		{\varphi _m}^k{\varphi _k}^n =  - \delta _m^n + {R^n}{\kappa _m}
		\end{array} \right.
	\end{equation}
	
	Given any contact structure, one can construct (many) geometric structures called almost contact structure by the procedure of polarization, although not all almost contact structure arise in this way. Given any almost contact structure, one can construct again many associated metric $g$ in the sense that
	\begin{equation}
		\kappa \left( R \right) = g\left( {R, \cdot } \right),\;\;{g_{mk}}{\varphi ^k}_n =  - {g_{nk}}{\varphi ^k}_m
	\end{equation}
The structure $(\kappa, R, \varphi, g)$ is called almost contact metric structure (ACMS).

	If an ACMS arises from some contact metric structure, such that
	\begin{equation}
		{\left( {d\kappa } \right)_{mn}} = {\varphi _{mn}}
	\end{equation}
then it is easy to see that
	\begin{equation}
		{\iota _R}*d\kappa  =  - d\kappa 
	\end{equation}
	
	The existence of contact structure in 5 dimension was not clear until very recently.  It is proved that every almost contact manifold admits contact structures, moreover, there is at least one contact structure within each homotopy class of almost contact structure\cite{Etnyre:2012aa}. There are also new results on distinguishing inequivalent contact structure as Boothby-Wang 5-manifolds \cite{Hamilton:2010aa}.

	Suppose $\nabla$ is any affine connection on $TM$, then one can define new connection $\hat \nabla$ that preserves $\varphi$:
	\begin{equation}
		{\hat \nabla _m}{X^n} \equiv {\nabla _m}{X^n} + {K^n}_{mk}{X^k}
	\end{equation}
where
	\begin{equation}
		{K^n}_{ml} \equiv  - \frac{1}{2}\left( {{\nabla _m}{\varphi _l}^k} \right){\varphi _k}^n - \frac{1}{2}{\kappa _l}{\nabla _m}{R^n} + {R^n}{\nabla _m}{\kappa _l}
	\end{equation}
If the affine connection $\nabla$ is chosen to be the Levi-civita connection associated to the ACMS structure, then one has
	\begin{equation}
		{K_{nml}} =  - {K_{lmn}}
	\end{equation}

	As mentioned, we have
	\begin{equation}
		{{\hat \nabla }_m}{\varphi _n}^k = 0
	\end{equation}
Moreover,  for any $X,Y \in \Gamma (TM_H)$, one has
	\begin{equation}
		g({\hat \nabla _X}Y , R) = 0
	\end{equation}
which means $\hat \nabla_X Y \in \Gamma(TM_H)$, the restriction of $\hat \nabla$ on $TM_H$ gives directly a connection $\hat \nabla |_{TM_H} \equiv \nabla^H$ on $TM_H$.
	
	The connection coefficients are now
	\begin{equation}
		{{{\hat \Gamma }}^n}\;_{ml} = {\Gamma ^n}_{ml} + {K^n}_{ml}
	\end{equation}
and the corresponding change of spin connection
	\begin{equation}
		\Delta {{ \omega }_{ma}}^b = {\omega _{ma}}^b + {K^b}_{ma}
	\end{equation}
where we define the spin connection\footnote{Note that the position of the flat indices $a$ and $b$ indicates that
	\begin{equation}
		{\nabla _m}\psi  = {\partial _m}\psi  - \frac{1}{4}{\omega _{mab}}{\Gamma ^{ab}}\psi 
	\end{equation}
as opposed to the frequently used notation ${{\omega _m}^b}_a$ which indicates
	\begin{equation}
		{\nabla _m}\psi  = {\partial _m}\psi  + \frac{1}{4}{\omega _{mab}}{\Gamma ^{ab}}\psi 
	\end{equation}
}
	\begin{equation}
		{\omega _{ma}}{^b} \equiv {e^b}\left( {{\nabla _m}{e_a}} \right) = {e^b}_n{\nabla _m}{e_a}^n = {e^b}_n{\partial _m}e_a^n + {\Gamma ^b}_{ma}
	\end{equation}

	In three dimension, where one has relation
	\begin{equation}
		{\varphi _{mn}} = {\epsilon _{mnk}}{R^k},\;\; R_m = \kappa_m
	\end{equation}
$K$ can be simplified as
	\begin{equation}
		{K^n}_{ml} = {R^n}{\nabla _m}{R_l} - {R_l}{\nabla _m}{R^n}
	\end{equation}
	
	The covariant derivative on spinor with new connection is now
	\begin{equation}
		{\hat \nabla _m}{\xi _I} = {\nabla _m}{\xi _I} - \frac{1}{4}{K_{lmn}}{\Gamma ^{nl}}\xi_I
	\end{equation}

	Now, let's consider the ACMS data coming from $\left( {{s^{ - 1}}R, s^{-1}\kappa ,r\left( t \right)t\Theta }, g \right)$,  where
	\begin{equation}
		r\left( t \right) = \frac{1}{s}\sqrt {\frac{{ - 2}}{{{\rm tr}\left( {{t^2}} \right)}}} 
	\end{equation}
such that
	\begin{equation}
		r{\left( t \right)^2}{\left( {t\Theta } \right)^2} =  - 1 + \left( {{s^{ - 2}}R} \right) \otimes \kappa 
	\end{equation}
	
	Substituting all these into definition of $K$, one has (with the assumption that $\Omega^- = 0$)
	\begin{equation}
		\begin{array}{l}
		\displaystyle {K_{nml}} = \frac{1}{{{s^2}}}\left( {{R_n}{\nabla _m}{R_l} - {R_l}{\nabla _m}{R_n}} \right) + \frac{1}{s}\frac{1}{{{\rm{tr}}\left( {{t^2}} \right)}}{\left( {{T_m}} \right)_{IJ}}{\left( {{\Theta ^{IJ}}} \right)_{ln}} \\[0.8em]
		\displaystyle\;\;\;\;\;\;\;\;\;\; - r{\left( t \right)^2}\left[ {{{\left( {*{V_V}} \right)}_{kmr}}{{\left( {t\Theta } \right)}_l}^r - {{\left( {*{V_V}} \right)}_{lmr}}{{\left( {t\Theta } \right)}_k}^r} \right]{\left( {t\Theta } \right)^k}_n
\end{array}
	\end{equation}	
where
	\begin{equation}
		{\left( {{T_m}} \right)_{IJ}} \equiv \left( {\nabla _m^A{t_I}^K} \right){t_{KJ}}
	\end{equation}
Note that when $s = 1$, $K_{nml} = - K_{lmn}$.
	
	To calculate the spin connection, one needs several convenient formula
	\begin{equation}
		{\left( {t\Theta } \right)_{nm}}{\Gamma ^n}{\xi _I} =  \left( {s{\Gamma _m} - {R_m}} \right){t_I}^J{\xi _J}
	\end{equation}
	\begin{equation}
		{\left( {t\Theta } \right)_{nm}} {\Gamma ^{kn}}{\xi _I}= {\Gamma ^k}\left( {{\Gamma _m}s - {R_m}} \right){t^J}_I{\xi _J} - {\left( {t\Theta } \right)^k}_m{\xi _I} \Rightarrow {\left( {t\Theta } \right)_{mn}}{\Gamma ^{mn}}{\xi _I} =  - 4s{t^J}_I{\xi _J}
	\end{equation}
	\begin{equation}
		{R^m}{\Gamma _{nm}}{\xi _I} = \left( {s{\Gamma _n} - {R_n}} \right){\xi _I}
	\end{equation}
	
	Finally, one has
	\begin{equation}
		\begin{array}{l}
		\displaystyle {{\hat \nabla }_m}{\xi _I} = {\nabla _m}{\xi _I} + \frac{1}{{{\rm{tr}}\left( {{t^2}} \right)}}{({T_m})^J}_I{\xi _J} - \frac{1}{{2s}}{\nabla _m}{R_n}{\Gamma ^n}{\xi _I} + \frac{1}{2}\left( {{\nabla _m}\log s} \right){\xi _I}\\[0.5em]
		\displaystyle\;\;\;\;\;\;\;\;\;\;\;\;\; - \frac{1}{{{\rm{tr}}\left( {{t^2}} \right)}}{\eta _q}{\left( {t\Theta } \right)^q}_m{t_I}^J{\xi _J} + \frac{1}{2}{\left( {*{V^V}} \right)_{mpq}}{\Gamma ^{pq}}{\xi _I}
		\end{array}
	\end{equation}
	
	Some remark. We used almost contact data $\varphi$ defined as $\sim t\Theta$, but in fact one could use any $SU(2)$ triplet function $\lambda$ to define $\varphi_\lambda \sim \lambda \Theta$, and in particular, one could choose $\lambda  = {\lambda _a}{\sigma ^a}$. It also has corresponding compatible connection $\hat \nabla_\lambda$, such that
	\begin{equation}
		\hat \nabla_\lambda {\varphi _\lambda } = 0
	\end{equation}
However, the tensor $K_{lmn}$ would not have the above simple form.

\section{Useful identities}\label{Appendix-E}
	\begin{equation}
		{\left| R \right|^2} = {R^m}{R_m} = {\iota _R}\kappa  = {s^2}
	\end{equation}
	\begin{equation}
		{\iota _R}{\Theta ^{IJ}} = 0
	\end{equation}
	\begin{equation}
		{\iota _R}*\Theta ^{IJ} = - s\Theta^{IJ} \Leftrightarrow {R^k}\left( {{\xi ^I}{\Gamma _{mnk}}{\xi ^J}} \right) =  + s\left( {{\xi ^I}{\Gamma _{mn}}{\xi ^J}} \right)
	\end{equation}
	\begin{equation}
		\kappa  \wedge \Theta  \wedge \Theta  \ne 0
	\end{equation}
	\begin{equation}
		{({\lambda _1}\Theta )_m}^p{({\lambda _2}\Theta )_p}^n = s{\left( {{\lambda _1}} \right)_I}^K{\left( {{\lambda _2}} \right)_{KJ}}{\left( {{\Theta ^{IJ}}} \right)_m}^n + \frac{{{s^2}}}{2}{\rm{tr}}\left( {{\lambda _1}{\lambda _2}} \right){\delta _m}^n - \frac{1}{2}{\rm{tr}}\left( {{\lambda _1}{\lambda _2}} \right){\kappa _m}{R^n}
	\end{equation}
	\begin{equation}
		{\left( {\lambda \Theta } \right)^{mn}}{\left( {\lambda \Theta } \right)_{mn}} = - 2s^2 {\rm tr}(\lambda^2)
		\label{2-Theta_1}
	\end{equation}
	\begin{equation}
		*{\left( {\lambda\Theta } \right)_{nkl}}{\left( {\lambda\Theta } \right)_m}^l =  \frac{s}{2}{\rm{tr}}\left( {{\lambda^2}} \right)\left[ {{g_{mk}}{R_n} - {g_{mn}}{R_k}} \right]
		\label{2-Theta_2}
	\end{equation}
	\begin{equation}
		{\left( {*\lambda\Theta } \right)^{mnk}}{\left( {\lambda\Theta } \right)_{mn}} =  2{\rm{tr}}\left( {{\lambda^2}} \right)s{R^k}
		\label{2-Theta_3}
	\end{equation}

	Also there are several useful spinor identities
	\begin{equation}
		{R^m}{\Gamma _m}{\xi _I} = s{\xi _I}
	\end{equation}
	\begin{equation}
		{R^m}{\Gamma _{nm}}{\xi _I} = \left( {s{\Gamma _n} - {R_n}} \right){\xi _I}
	\end{equation}
	\begin{equation}
		{\left( {\lambda \Theta } \right)_{nm}}{\Gamma ^n}{\xi _I} = \left( {{R_m} - s{\Gamma _m}} \right){\lambda_I}^J{\xi _J}
	\end{equation}
	\begin{equation}
		{\left( {\lambda \Theta } \right)_{nm}}{\Gamma ^{kn}}{\xi _I} = {\Gamma ^k}\left( {{R_m} - s{\Gamma _m}} \right){\lambda ^J}_I{\xi _J} - {\left( {\lambda \Theta } \right)^k}_m{\xi _I} \Rightarrow {\left( {\lambda \Theta } \right)_{mn}}{\Gamma ^{mn}}{\xi _I} = 4s{\lambda^J}_I{\xi _J}
	\end{equation}

\section{Proof of Supersymmetry Invariance}\label{Appenidx-F}

	Let us focus on the part of Lagrangian (with trace left implicit) without the "topological term" $\kappa\wedge F\wedge F$ and the scalar curvature term:
	\begin{equation}
		\begin{array}{l}
		\displaystyle \mathcal{L} = \frac{1}{2}{F_{mn}}{F^{mn}} - {D_m}\sigma {D^m}\sigma  - \frac{1}{2}{D_{IJ}}{D^{IJ}} + i{\lambda _I}{\Gamma ^m}{D_m}{\lambda ^I} - {\lambda _I}\left[ {\sigma ,{\lambda ^I}} \right]\\[0.5em]
		\;\;\;\;\;\;\;\displaystyle - i{t^{IJ}}\left( {{\lambda _I}{\lambda _J}} \right) + 2\sigma {t^{IJ}}{D_{IJ}} + \frac{i}{4}{{{\cal F}}_{mn}}\left( {{\lambda _I}{\Gamma ^{mn}}{\lambda ^I}} \right) + {{{\cal F}}^{mn}}{F_{mn}}\sigma 
		\end{array}
	\end{equation}
	
	Its supersymmetry variation (partial integration has been used for $\lambda$ kinetic term)
	\begin{equation}
		\begin{array}{l}
		\delta \mathcal{L}= \displaystyle {F^{mn}}\delta {F_{mn}} - 2{D_m}\sigma \delta \left( {{D^m}\sigma } \right) - {D^{IJ}}\delta {D_{IJ}} + 2i\delta {\lambda _I}{\Gamma ^m}{D_m}{\lambda ^I} + {\lambda _I}{\Gamma ^m}\left[ {\delta {A_m},{\lambda ^I}} \right] \\[0.8em]
		\;\;\;\;\;\;\;\;\;\;\displaystyle - \left\{ {2\delta {\lambda _I}\left[ {\sigma ,{\lambda ^I}} \right] + {\lambda _I}\left[ {\delta \sigma ,{\lambda ^I}} \right]} \right\}- 2i{t^{IJ}}\left( {\delta {\lambda _I}{\lambda _J}} \right) + 2\left\{ {\delta \sigma {t^{IJ}}{D_{IJ}} + \sigma {t^{IJ}}\delta {D_{IJ}}} \right\}\\[0.8em]
		 \displaystyle\;\;\;\;\;\;\;\;\;\;+ \frac{i}{2}{{{\cal F}}_{mn}}\left( {\delta {\lambda _I}{\Gamma ^{mn}}{\lambda ^I}} \right)+ \left\{ {{{{\cal F}}^{mn}}\delta {F_{mn}}\sigma  + {{{\cal F}}^{mn}}{F_{mn}}\delta \sigma } \right\}
		\end{array}
	\end{equation}
	
	I) Let us simplify the first row (contributed by flat SUSY Lagrangian) denoted by \textbf{I}.
	\begin{equation}
		\begin{array}{l}
		\textbf{I} = \displaystyle {\color{blue}2{F^{mn}}{D_m}\left( {i{\xi _I}{\Gamma _n}{\lambda ^I}} \right)} - 2{D^m}\sigma \left\{ {\color{green}{{D_m}\left( {i{\xi _I}{\lambda ^I}} \right)} +{\color{red} \left[ {\left( {{\xi _I}{\Gamma _m}{\lambda ^I}} \right),\sigma } \right]}} \right\}\\[0.8em]
		\displaystyle\;\;\;\;\;\; - 2{D^{IJ}}\left( {\color{red}{ - i{\xi _I}{\Gamma ^m}{D_m}{\lambda _J}} + {\color{red}\left[ {\sigma ,\left( {{\xi _I}{\lambda _J}} \right)} \right]}} + i({{\tilde \xi }_I}{\lambda _J}) \right)\\[0.8em]
		\displaystyle\;\;\;\;\;\; + 2i\left( { {\color{blue}- \frac{1}{2}{F_{pq}}{\Gamma ^{pq}}{\xi _I}} +{\color{green} \left( {{D_n}\sigma } \right){\Gamma ^n}{\xi _I}} + {\color{red}{D_I}^J{\xi _J}} + 2\sigma {{\tilde \xi }_I}} \right){\Gamma ^m}{D_m}{\lambda ^I}+ {\color{red}i{\lambda _I}{\Gamma ^m}\left[ {\left( {{\xi _J}{\Gamma _m}{\lambda ^J}} \right),{\lambda ^I}} \right]}\\[0.8em]
		\displaystyle\;\;\; \;\;\;- \left\{ {2\left( { {\color{green}- \frac{1}{2}{F_{pq}}{\Gamma ^{pq}}{\xi _I}} + {\color{red} \left( {{D_n}\sigma } \right){\Gamma ^n}{\xi _I}} + {\color{red}{D_I}^J{\xi _J}} + 2\sigma {{\tilde \xi }_I}} \right)\left[ {\sigma ,{\lambda ^I}} \right] +{\color{red} i{\lambda _I}\left[ {\left( {{\xi _J}{\lambda ^J}} \right),{\lambda ^I}} \right]}} \right\}
		\end{array}
	\end{equation}
	
	(1) Immediate cancelation between the {\color{red} red} terms.
	
	(2) The {\color{blue} blue} terms  add up to
	\begin{equation}
		\begin{array}{l}
		\displaystyle 2{F^{mn}}{D_m}\left( {i{\xi _I}{\Gamma _n}{\lambda ^I}} \right) + \left( {2i} \right)\left( {\frac{1}{2}} \right){F_{pq}}\left( {{\xi _I}{\Gamma ^{pq}}{\Gamma ^m}{D_m}{\lambda ^I}} \right)\\[0.8em]
		 = 2i{F^{mn}}( {{{\tilde \xi }_I}{\Gamma _{mn}}{\lambda ^I}} ) + i{F_{pq}}\left( {{\xi _I}{\Gamma ^{pqm}}{D_m}{\lambda ^I}} \right)\\[0.8em]
		 =  {\color{blue}- i{F_{mn}}( {{{\tilde \xi }_I}{\Gamma _{mn}}{\lambda ^I}} )}
		\end{array}
	\end{equation}
	
	(3) The {\color{green} green} terms add up to
	\begin{equation}
		\begin{array}{l}
		 - 2{D^m}\sigma {D_m}\left( {i{\xi _I}{\lambda ^I}} \right) + 2i\left( {{D_n}\sigma } \right)\left( {{\xi _I}{\Gamma ^n}{\Gamma ^m}{D_m}{\lambda ^I}} \right) - {F_{pq}}{\xi _I}{\Gamma ^{pq}}\left[ {\sigma ,{\lambda ^I}} \right]\\[0.8em]
		\displaystyle  ={\color{green}  - 2{D^m}\sigma ( {i{{\tilde \xi }_I}{\Gamma _m}{\lambda ^I}} ) - 8i\sigma ( {{{\tilde \xi }_I}{\Gamma ^m}{D_m}{\lambda ^I}} ) - {\delta _\xi }\left( {\frac{1}{4}{{\cal R}}{\sigma ^2}} \right) }
		\end{array}
	\end{equation}
where the last term cancels the scalar curvature term in the Lagrangian\footnote{Note that there are two ways of doing partial integration for the second term: one gives $\left[ {{D_m},{D_n}} \right]\sigma \left( {{\xi _I}{\Gamma ^{nm}}{\lambda ^I}} \right)$ while the other $\sigma \left( {{\xi _I}{\Gamma ^{nm}}\left[ {{D_m},{D_n}} \right]{\lambda ^I}} \right)$. The former way directly produces ${\rm{tr}}\left( {{t_{IJ}}{t^{IJ}}{\sigma ^2}} \right)$ which appears in \cite{Hosomichi:2012fk}, but the cancellation of other terms are relatively tricky. Hence we take the second way of doing partial integration, which gives instead ${\rm tr}(\mathcal{R} \sigma^2)$}.
	
	(4) Now, we can gather all the terms and obtain the leftovers
	\begin{equation}
		\textbf{I} = {\color{red}- 2i{D^{IJ}}({\tilde \xi _I}{\lambda _J})} - 4i\sigma ({\tilde \xi _I}{\Gamma ^m}{D_m}{\lambda ^I}) {\color{blue}- i{F_{mn}}({\tilde \xi _I}{\Gamma _{mn}}{\lambda ^I})} - 2i\left( {{D^m}\sigma } \right)({\tilde \xi _I}{\Gamma _m}{\lambda ^I})
	\end{equation}

	II) Now we try to simplify the 2nd and 3rd row, and denote it by \textbf{II}. 
	\begin{equation}
		\begin{array}{l}
		\;\;\;\;\;\;\;\;\;\;\displaystyle - \left\{ {2\delta {\lambda _I}\left[ {\sigma ,{\lambda ^I}} \right] + {\lambda _I}\left[ {\delta \sigma ,{\lambda ^I}} \right]} \right\}- 2i{t^{IJ}}\left( {\delta {\lambda _I}{\lambda _J}} \right) + 2\left\{ {\delta \sigma {t^{IJ}}{D_{IJ}} + \sigma {t^{IJ}}\delta {D_{IJ}}} \right\}\\[0.8em]
		 \displaystyle\;\;\;\;\;\;\;\;\;\;+ \frac{i}{2}{{{\cal F}}_{mn}}\left( {\delta {\lambda _I}{\Gamma ^{mn}}{\lambda ^I}} \right)+ \left\{ {{{{\cal F}}^{mn}}\delta {F_{mn}}\sigma  + {{{\cal F}}^{mn}}{F_{mn}}\delta \sigma } \right\}
		\end{array}
	\end{equation}
Explicitly writing down all terms,
	\begin{equation}
		\begin{array}{l}
		\displaystyle \textbf{II}=  - 2i{t^{IJ}}\left(  {{\color{blue}\frac{1}{2}{F_{pq}}{\xi _I}{\Gamma ^{pq}}{\lambda _J}} + \left( {{D_p}\sigma } \right){\xi _I}{\Gamma ^p}{\lambda _J} + {\color{red}{D_I}^K{\xi _K}{\lambda _J}} + {\color{red}2\sigma {{\tilde \xi }_I}{\lambda _J}}}  \right)\\[0.5em]
		\;\;\;\;\;\;\;\; + {\color{red}2\left( {i{\xi _K}{\lambda ^K}} \right){t^{IJ}}{D_{IJ}}}\\[0.5em]
		\;\;\;\;\;\;\;\; - 4i\sigma {t^{IJ}}\left( {{\xi _I}{\Gamma ^m}{D_m}{\lambda _J}} \right) + {\color{red}4i\sigma {t^{IJ}}( {{{\tilde \xi }_I}{\lambda _J}} )}\\[0.5em]
		\displaystyle\;\;\;\;\;\;\;\; + \frac{i}{2}{{{\cal F}}_{mn}}\left( {{\color{blue}\frac{1}{2}{F_{pq}}\left( {{\xi _I}{\Gamma ^{pq}}{\Gamma ^{mn}}{\lambda ^I}} \right)} + \left( {{D_p}\sigma } \right){\xi _I}{\Gamma ^p}{\Gamma ^{mn}}{\lambda ^I} + \boxed{\color{red}{D_I}^K{\xi _K}{\Gamma ^{mn}}{\lambda ^I}} + 2\sigma {{\tilde \xi }_I}{\Gamma ^{mn}}{\lambda ^I}} \right)\\[0.5em]
		\;\;\;\;\;\;\;\; + {\color{blue}{{{\cal F}}^{mn}}{F_{mn}}\left( {i{\xi _I}{\lambda ^I}} \right)}\\[0.5em]
		\;\;\;\;\;\;\;\; + 2i{{{\cal F}}^{mn}}\sigma {D_m}\left( {{\xi _I}{\Gamma _n}{\lambda ^I}} \right)
		\end{array}
	\end{equation}
	
	Combining with leftover of \textbf{I}, one sees
	
	(1) Some immediately cancels in {\color{red} red}
	\begin{equation}
		2i{t^{IJ}}{D_{IJ}}\left( {{\xi _K}{\lambda ^K}} \right) - 2i{t^{IJ}}{D_I}^K\left( {{\xi _K}{\lambda _J}} \right) + 2i{D^{IJ}}\left( {{{\tilde \xi }_I}{\lambda _J}} \right) = {\color{red}\frac{i}{2}{{{\cal F}}_{mn}}{D^{IJ}}\left( {{\xi _I}{\Gamma ^{mn}}{\lambda _J}} \right)}
	\end{equation}
where one needs identity
	\begin{equation}
		{t^J}_K\left( {{\xi ^I}{\lambda ^K}} \right) + {t^I}_K\left( {{\xi ^J}{\lambda ^K}} \right) + 2{t^{IJ}}\left( {{\xi ^K}{\lambda _K}} \right) = {t_K}^I\left( {{\xi ^K}{\lambda ^J}} \right) + {t_K}^J\left( {{\xi ^K}{\lambda ^I}} \right)
	\end{equation}
The leftover $\mathcal{F} D \lambda$ terms cancels the \boxed{\color{red}boxed} term
	\begin{equation}
		\frac{i}{2}{{{\cal F}}_{mn}}{D_I}^K\left( {{\xi _K}{\Gamma ^{mn}}{\lambda ^I}} \right) + \frac{i}{2}{F_{mn}}{D^{IJ}}\left( {{\xi _K}{\Gamma ^{mn}}{\lambda _I}} \right) = 0
	\end{equation}

	(2) The $tF\lambda$ terms and $F\mathcal{F} \lambda$ terms in {\color{blue} blue} add up to
	\begin{equation}
		\begin{array}{l}
		\displaystyle - 2i{t^{IJ}}\frac{1}{2}{F_{pq}}\left( {{\xi _I}{\Gamma ^{pq}}{\lambda _J}} \right) - i{F^{mn}}( {{{\tilde \xi }_I}{\Gamma _{mn}}{\lambda ^I}} ) + \frac{i}{4}{{{\cal F}}_{mn}}{F_{pq}}\left( {{\xi _I}{\Gamma ^{pq}}{\Gamma ^{mn}}{\lambda ^I}} \right) + {{{\cal F}}^{mn}}{F_{mn}}\left( {i{\xi _I}{\lambda ^I}} \right)\\[0.8em]
		 \displaystyle= \frac{i}{4}{{{\cal F}}_{mn}}{F_{pq}}\left( {{\xi _I}{\Gamma ^{mn}}{\Gamma ^{pq}}{\lambda ^I}} \right) + \frac{i}{4}{{{\cal F}}_{mn}}{F_{pq}}\left( {{\xi _I}{\Gamma ^{pq}}{\Gamma ^{mn}}{\lambda ^I}} \right) + {{{\cal F}}^{mn}}{F_{mn}}\left( {i{\xi _I}{\lambda ^I}} \right)\\[0.8em]
		 \displaystyle= \frac{i}{4}{{{\cal F}}_{mn}}{F_{pq}}\left( {{\xi _I}\left\{ {{\Gamma ^{mn}},{\Gamma ^{pq}}} \right\}{\lambda ^I}} \right) + {{{\cal F}}^{mn}}{F_{mn}}\left( {i{\xi _I}{\lambda ^I}} \right)\\[0.8em]
		 \displaystyle= {\color{blue}\frac{i}{2}{{{\cal F}}_{mn}}{F_{pq}}\left( {{\xi _I}{\Gamma ^{mnpq}}{\lambda ^I}} \right)}
		\end{array}
	\end{equation}
where one needs identity
	\begin{equation}
		\left\{ {{\Gamma ^{mn}},{\Gamma ^{pq}}} \right\} = 2\left( {{g^{np}}{g^{mq}} - {g^{mp}}{g^{nq}}} \right) + 2{\Gamma ^{mnpq}}
	\end{equation}
	
	The term remaining can actually be written as
	\begin{equation}
		\frac{i}{2}{{{\cal F}}_{mn}}{F_{pq}}\left( {{\xi _I}{\Gamma ^{mnpq}}{\lambda ^I}} \right)\sqrt g {d^5}x = \delta \left( {\frac{{\sqrt g }}{4}{\epsilon ^{mnpqr}}{\kappa _r}{F_{pq}}{F_{mn}}\sqrt g {d^5}x} \right) = \delta \left( {\kappa  \wedge F \wedge F} \right)
	\end{equation}
This term cancels the "topological" terms in the proposed action.

	(3) The remaining black terms reads
	\begin{equation}
		\begin{array}{l}
		 \displaystyle- 2i{t^{IJ}}\left( {{D_p}\sigma } \right)\left( {{\xi _I}{\Gamma ^p}{\lambda _J}} \right) - 4i\sigma {t^{IJ}}\left( {{\xi _I}{\Gamma ^m}{D_m}{\lambda _J}} \right) + \frac{i}{2}{{{\cal F}}_{mn}}\left( {{D_p}\sigma } \right)\left( {{\xi _I}{\Gamma ^p}{\Gamma ^{mn}}{\lambda ^I}} \right)\\[0.5em]
		 \displaystyle+ \frac{i}{2}{{{\cal F}}_{mn}}2\sigma ( {{{\tilde \xi }_I}{\Gamma ^{mn}}{\lambda ^I}} ) + 2i{{{\cal F}}^{mn}}\sigma {D_m}\left( {{\xi _I}{\Gamma _n}{\lambda ^I}} \right) - 2i\left( {{D^m}\sigma } \right)( {{{\tilde \xi }_I}{\Gamma _m}{\lambda ^I}} ) - 4i\sigma ( {{{\tilde \xi }_I}{\Gamma ^m}{D_m}{\lambda ^I}} )
		\end{array}
	\end{equation}
Explicit computation shows these terms cancel each other.

\end{appendices}

\bibliographystyle{JHEP}
\bibliography{5d_SUSY}

\end{document}